\documentclass{vgtc}                          

\ifpdf%
  \pdfoutput=1\relax                   
  \usepackage{graphicx}               
  \DeclareGraphicsExtensions{.pdf,.png,.jpg,.jpeg} 
\else%
  \usepackage{graphicx}              
  \DeclareGraphicsExtensions{.eps}     
\fi%

\graphicspath{{figures/}{pictures/}{images/}{./}} 

\usepackage{microtype}                 
\PassOptionsToPackage{warn}{textcomp}  
\usepackage{textcomp}                  
\usepackage{mathptmx}                  
\usepackage{amsfonts}
\usepackage{times}                     
         
\usepackage{cite}                      
\usepackage{tabu}  
\usepackage{subcaption}
\usepackage{booktabs}                  
\usepackage{makecell}
\usepackage{xspace}
\usepackage{enumitem}
\usepackage{amsmath}
\usepackage{subfiles}
\usepackage{overpic}
\usepackage{ccicons}
\usepackage{textcomp}
\usepackage{multirow}
\usepackage{makecell}

\usepackage{makecell} 
\usepackage{cuted}
\usepackage{xcolor}
\usepackage[ruled,vlined,linesnumbered]{algorithm2e}
\usepackage{tikz}

\usepackage[normalem]{ulem} 
\SetKwProg{Fn}{function}{}{} 
\SetKwFunction{PDE}{PDE}
\SetKwFunction{ASL}{ASL}
\SetKwFunction{FDE}{FDE}
\SetKwFunction{FSKDE}{FSKDE}
\DontPrintSemicolon                    
\SetKwInOut{Input}{Input}
\SetKwInOut{Output}{Output}
\SetKwProg{Fn}{function}{:}{end}     
\SetKw{KwTo}{to}
\SetKwIF{If}{ElseIf}{Else}{if}{then}{else if}{else}{end}
\SetKwFor{For}{for}{do}{end}
\SetKwFor{ForEach}{for each}{do}{end}
\SetInd{0em}{0.7em}                     
\SetAlgoNlRelativeSize{-1}             
\SetAlgoCaptionSeparator{:\ }        
\SetCommentSty{textcolor{gray}{footnotesize}}

\newcommand{\etal}{et~al.\xspace}
\newcommand{\eg}{e.\,g.}
\newcommand{\ie}{i.\,e.}

\newcommand{\method}{ScaleFree\xspace}

\newcommand{\FigSubref}[2]{%
  \hyperref[#2]{\autoref*{#1}(\subref*{#2})}%
}

\definecolor{brandeisblue}{rgb}{0.063, 0.506, 0.78}

\newcommand{\lixiang}[1]{\textcolor{black}{#1}}
\newcommand{\lingyun}[1]{\textcolor{black}{#1}}

\newlength{\lineskiplimitbackup}%
\newlength{\lineskipbackup}%
\newlength{\mathskip}%
\newcommand{\compactmathparnominipage}[1]{%
	\setlength{\lineskiplimit}{-\maxdimen}%
	\setlength{\lineskip}{0pt}%
  #1%
	}
\newcommand{\compactmathpar}[1]{%
	\begin{minipage}{\linewidth}%
	\compactmathparnominipage{#1}%
	\end{minipage}%
	}

\newcommand{\backuplineskips}{%
	\setlength{\lineskiplimitbackup}{\lineskiplimit}%
	\setlength{\lineskipbackup}{\lineskip}%
}
\newcommand{\restorelineskips}{%
	\setlength{\lineskiplimit}{\lineskiplimitbackup}%
	\setlength{\lineskip}{\lineskipbackup}%
}
\newcommand{\extranegmathskip}{\vspace{-\mathskip}}%
\newcommand{\mathcompact}[1]{\raisebox{0pt}[0pt][0pt]{#1}}

\onlineid{1680}

\vgtccategory{Research}

\vgtcinsertpkg

\title{\method: Dynamic KDE for Multiscale Point Cloud Exploration in VR}

\author{
Lixiang Zhao$^{1}$\,\href{https://orcid.org/0000-0001-6181-1673}{\includegraphics[height=1.5ex]{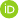}} \and
Fuqi Xie$^{1}$\,\href{https://orcid.org/0009-0008-4728-9346}{\includegraphics[height=1.5ex]{orcid}} \and
Tobias Isenberg$^{2}$\,\href{https://orcid.org/0000-0001-7953-8644}{\includegraphics[height=1.5ex]{orcid}} \and
Hai-Ning Liang$^{3}$\,\href{https://orcid.org/0000-0003-3600-8955}{\includegraphics[height=1.5ex]{orcid}} \and
Lingyun Yu$^{1}$\thanks{
Corresponding author: Lingyun Yu (\href{mailto:lingyun.yu@xjtlu.edu.cn}{lingyun.yu@xjtlu.edu.cn}).} \,\href{https://orcid.org/0000-0002-3152-2587}{\includegraphics[height=1.5ex]{orcid}
}
}

\affiliation{\scriptsize
$^{1}$ Xi'an Jiaotong-Liverpool University, Suzhou, China\hspace{10mm}
$^{2}$ Universit{\'e} Paris-Saclay, CNRS, Inria, LISN, France\\
$^{3}$ Hong Kong University of Science and Technology (Guangzhou), China
}

\teaser{
  \centering
  \includegraphics[width=.8\linewidth]{figures/Teaser.png}
  \caption{\method dynamically recomputes density fields on the GPU, enabling on-the-fly density estimation of regions of interest (ROI) at any scale and position within massive multiscale point clouds in VR.}
  \label{fig:teaser}
}

\abstract{
We present \method, a GPU-accelerated adaptive Kernel Density Estimation (KDE) algorithm for scalable, interactive multiscale point cloud exploration. With this technique, we cater to the massive datasets and complex multiscale structures in advanced scientific computing, 
such as cosmological simulations with billions of particles. Effective exploration of such data requires a full 3D understanding of spatial structures, a capability for which immersive environments such as VR are particularly well suited. However, simultaneously supporting global multiscale context and fine-grained local detail remains a significant challenge.
A key difficulty lies in dynamically generating continuous density fields from point clouds to facilitate the seamless scale transitions: while KDE is widely used, precomputed fields restrict the accuracy of interaction and omit fine-scale structures, while dynamic computation is often too costly for real-time VR interaction. We address this challenge by leveraging GPU acceleration with k-d-tree-based spatial queries and parallel reduction within a thread group for on-the-fly density estimation. With this approach, we can recalculate scalar fields dynamically as users shift their focus across scales. We demonstrate the benefits of adaptive density estimation through two data exploration tasks: adaptive selection and progressive navigation. Through performance experiments, we demonstrate that \method with GPU-parallel implementation achieves orders-of-magnitude speedups over sequential and multi-core CPU baselines. In a controlled experiment, we further confirm that our adaptive selection technique improves accuracy and efficiency in multiscale selection tasks.} 

\keywords{Multiscale visualization, kernel density estimation, interactive data selection, interactive data navigation, VR/AR/MR.}

\begin{document}



\maketitle
\section{Introduction}
\label{sec:Introduction}
Advances in scientific computing have produced datasets that are not only larger in size but also richer in multiscale structural complexity. 
For example, modern cosmological simulations produce billions of particles, with details that \lixiang{span several} orders of magnitude, to capture the formation and evolution of cosmic structures \cite{Springel:2005:SimulationsOT,springel:2008:aquarius}.
The fact that meaningful features often emerge at different scales requires analysts to fluidly transition between global and local perspectives---for instance, from cosmic filaments and clusters to fine-grained substructures in astronomical point clouds~\cite{Springel:2005:SimulationsOT,hahn:2007:PDM}. 
To understand the dynamics of structure formation, scientists need to visualize and explore the three-dimensional spatial patterns across multiple scales.
Immersive environments delivered through virtual, augmented, and mixed reality head-mounted displays (HMDs) have emerged as superior tools for interpreting complex spatial relationships. 
\lingyun{Prior research has shown that immersive visualizations can enhance users' understanding of complex spatial structures, including volume data \cite{laha:2012:EOI}, astronomical data \cite{zhao:2022:Lwim}, and node-link diagrams \cite{garcia:2016:PFU,whitlock:2020:GPF}. These benefits are particularly evident in navigation \cite{sousa:2009:HMD} and 3D manipulation tasks \cite{bueckle:2021:3vr}, where frequent viewpoint changes are required to inspect spatial structures.}
The immersive environment not only provide stereoscopic perception of 3D structures but also support seamless transitions between global overviews and localized detail exploration. For instance, a user may zoom in to inspect a galaxy cluster and then zoom out to contextualize its dynamics within the surrounding large-scale structures. This process requires not only an immersive environment, such as VR, but also high-performance scientific computing and visualization systems capable of supporting free navigation across space, time, and scale.

Achieving smooth transitions and effective exploration across multiple scales remains challenging, however, in particular when data must be transformed into an alternative representation to support further analysis. 
For example, KDE~\cite{diggle:1985:KDEDiggle} has been widely used in point cloud visualization to uncover clusters and structural patterns in large multiscale datasets by converting discrete points into a continuous scalar density field. 
Such density fields are fundamental to spatial interaction, as they provide a robust basis for feature detection \cite{chen:2015:CWR,pfeifer:2022:COWS,shivashankar:2015:Felix,cautun:2013:NEXUS} and for constructing selection volumes~\cite{Yu:2016:CEE,Yu:2012:ESA,zhao:2023:metacast}, thereby enabling users to identify, isolate, and manipulate regions of interest.
Nonetheless, computing scalar density fields on the fly is computationally expensive and often fails to guarantee real-time interaction. \lingyun{Prior density-based selection techniques for point cloud exploration~\cite{Yu:2012:ESA,Yu:2016:CEE,zhao:2023:metacast} have typically focused on selection interaction itself and relied on precomputed, single-scale density fields, rather than targeting dynamic multiscale recomputation during interaction.}
Such precomputation, however, implies that the density fields are static during exploration, constraining selection accuracy to the resolution defined \emph{a priori} and leaving finer-scale structures unrepresented. 
\lingyun{This limitation becomes particularly pronounced in VR. Unlike desktop systems, VR presents users with a single, continuous egocentric view of the data space, where navigation, selection, and scale transitions are tightly coupled with head and body movement. Users frequently shift their focus between global overviews and localized regions, creating a strong demand for density fields that adapt immediately to scale and perspective changes. At the same time, VR imposes strict real-time constraints due to high-frequency stereoscopic rendering and low-latency interaction requirements. These challenges motivate us to explore 
density estimation techniques that can be dynamically computed at varying scales during real-time immersive exploration.}

We thus introduce \method, a fast, adaptive, and scalable KDE algorithm specifically optimized for interactive point cloud data analysis. 
Our method leverages GPU acceleration to perform on-the-fly density estimation, enabling scalar fields to be recomputed dynamically as users shift their focus across scales. By maintaining density fields that adapt repeatedly
rather than relying on precomputed representations, \method facilitates more precise and responsive support for core exploration tasks such as \emph{spatial selection}, \emph{feature localization}, and \emph{multiscale navigation} in large point cloud datasets. 
\lingyun{To evaluate the role of adaptive KDE in interactive VR exploration, we adopt a multi-layered evaluation strategy. We first compare our GPU-based implementation against CPU baselines to establish real-time feasibility, and then evaluate how integrating adaptive KDE affects selection effectiveness through technique demonstrations and a controlled user study, measuring accuracy, efficiency, and workload against state-of-the-art VR selection techniques.}
Specifically, we conducted a user study with 24 participants comparing three \emph{selection techniques} that employ different density estimation approaches, including precomputed single-resolution density fields (PS), precomputed multi-resolution mipmap density fields (PM), and our dynamic \method approach. 
Results show that selections made with \method were faster and more accurate than those using precomputed density fields. Participants also reported lower workload and expressed a stronger preference for \method when conducting multiscale exploration tasks in VR. 
\lingyun{In summary, our GPU-accelerated adaptive KDE makes the following main contributions to large-scale spatial data exploration in immersive environments:
\begin{itemize}[nosep]
\item \textbf{Adaptive and smooth scale transitions:} enabling continuous transitions across scales without disruptive visual artefacts or noticeable delay (demonstrated in both selection \autoref{subsec:selection:method} and navigation \autoref{subsec:navigation:method} scenarios).
\item \textbf{Stable viewpoint-driven exploration:} supporting smooth, orientation-preserving transitions between overview and localized perspectives across scales during exploration of complex datasets (demonstrated in the navigation scenario, \autoref{subsec:navigation:method}).
\item \textbf{Fast and responsive computation:} improving selection accuracy and efficiency under varying scales, thereby supporting precise spatial interaction in immersive analytics (shown in the user study results, \autoref{sec:userstudy}).
\end{itemize}}

\section{Related work}
\label{sec:related}
Our work builds on prior research in multiscale immersive exploration and density estimation methods that support such exploration.

\subsection{Multiscale Exploration in Immersive Environment}
Multiscale exploration requires users to operate across different levels of scale, ideally without relying on explicit mode switches or commands. Prior studies have demonstrated the advantages of immersive environments in diverse scientific domains such as astronomy~\cite{Song:1994:LIL, Fu:2010:MTT}, geography~\cite{Cho:2018:MS7}, biology~\cite{Kouvril:2021:HyperLabels}, and architecture~\cite{argelaguet:2016:giant, LaViola:2001:HFM}. A widely adopted technique for multiscale navigation in immersive environments relies on a \emph{World-in-Miniature} (WIM) \cite{stoakley:1995:VRWIM}, which presents both the life-size virtual environment and a hand-held, scaled-down replica that functions as an additional viewport. This dual representation enables users to interact with the environment at multiple scales: they can directly manipulate objects via the miniature, specify regions of interest for navigation, and then seamlessly ``jump'' to the corresponding areas in the full-scale world.
Subsequent variations of WIM, such as Scalable WIM~\cite{pivovar:2022:SWIM} and Scaled and Scrolling WIM~\cite{wingrave:2006:SSWIM}, extended the metaphor by introducing panning and rescaling of the miniature, facilitating navigation across broader spatial ranges and more precise exploration of fine-grained structures, overcoming the scale and resolution constraints of the original WIM. 
Similarly, \emph{target-based} and \emph{steering-based} approaches, such as the magnifying glass metaphor by Kopper~\etal~\cite{kopper:2006:DEN}, enable navigation through predefined discrete scale levels.
Although these methods provide overview-detail support, however, they often fail to preserve users' sense of scale when ``jumping'' across scale levels. 

To address this issue, Zhao~\etal~\cite{zhao:2022:Lwim} linked multiple WIMs into a \emph{hierarchical structure}, allowing users to seamlessly ``jump'' between different scales. Similarly, Bacim~\etal~\cite{bacim:2009:WTM} incorporated a hierarchical map to aid travel and wayfinding in anatomy applications. Moving beyond hierarchical structures, \emph{label-based} approaches employ meaningful anchors that reflect data organization, as in Kou{\v{r}}il~\etal's \cite{Kouvril:2021:HyperLabels} approach, who used active labels to browse hierarchical molecular visualizations. 
Nevertheless, these methods all depend on well-defined targets and hierarchical algorithms to compute and represent scale levels, with transitions typically occurring in either an end-to-end or step-by-step manner. Similar transition approaches are found in molecular visualization~\cite{Halladjian:2022:MUI, Halladjian:2020:ScaleTrotter}, where hierarchical molecular structures provide clearly defined scale levels. By contrast, we focus on unstructured and massive scientific datasets, such as astronomical point clouds, where predefined or meaningful hierarchies are often absent. In such scenarios, ``jumping'' approaches are less straightforward---we need more flexible techniques that support navigation and interaction at arbitrary scales.

\subsection{\lixiang{Key Aspects in Multiscale Immersive Exploration}}
\lingyun{Building on prior work in multiscale immersive exploration, we next synthesize key design aspects and limitations identified across existing systems. In particular, we focus on challenges related to fast and scalable computation, which motivate the need for adaptive density estimation techniques reviewed in the following subsection.}

\textbf{Scalable and flexible exploration.} 
Effective multiscale exploration should enable users to select or navigate any arbitrary region within an object or space, as features may appear at any scale. When hierarchical information is not explicitly available, such as in astronomical point cloud data~\cite{Springel:2005:SimulationsOT,hahn:2007:PDM} or additive manufacturing objects~\cite{Chheang:2024:AVE, Klacansky:2022:VIA} with dense, homogeneous internal structures, localization and navigation become particularly challenging. In such cases,  continuous zooming enables more flexible and fluid multiscale exploration. To address this issue, Pavanatto~\etal~\cite{pavanatto:2025:EMN} proposed a progressive refinement approach, facilitating, step-by-step, focused inspections via a selection box. Our work shares this motivation but takes a different path: rather than refining predefined regions, we enable users to focus on relatively dense areas across diverse scales. This approach, in turn, requires a dynamic computation of density fields to adaptively support the exploration.

\textbf{Adaptive and smooth transition.}  
In free exploration, adaptive transitions are essential to reduce disorientation and preserve a natural sense of immersion in multiscale environments. To ensure continuity, techniques should dynamically adjust factors such as computational scale, navigation speed~\cite{Colin:1997:CSF, argelaguet:2016:giant}, and the visualization's and users' relative scale~\cite{argelaguet:2016:giant, weissker:2024:TTF}. Traversing multiscale data, for instance, often requires adaptive ``flying'' speeds to balance comfort and continuity. Ware and Fleet~\cite{Colin:1997:CSF} addressed this point by introducing continuous depth sampling to modulate the navigation speed, while Argelaguet~\etal~\cite{argelaguet:2016:giant} extended the concept with an adaptive navigation technique that facilitates direct camera control, while it also automatically adjusts speed and environmental scale based on object distance and optical flow. More recently, Weissker~\etal~\cite{weissker:2024:TTF} proposed \emph{teleportation-based} multiscale travel methods that integrate scale adjustments directly into the teleportation process. 
All this work highlights the need for approaches that enable users to transition smoothly through unstructured multiscale data, with navigation, visualization, and interaction parameters continuously adapted to the new scale, all of which we also strive for in our work.

\textbf{Fast and responsive computation.}  
Finally, achieving these interaction designs in practice requires scalable computational support: Techniques must remain efficient and responsive when applied to large, complex datasets, ensuring real-time performance. One common approach is to introduce hierarchies of reduced objects or resolutions to balance rendering and interaction costs. Gansner~\etal~\cite{Gansner:2005:TFV}, for example, precomputed a hierarchy of coarsened graphs that could be combined on-the-fly for visualization. For massive unstructured datasets such as point clouds, density estimation methods like KDE form the foundation for feature detection, region-of-interest selection, and navigation across scales. The key challenge, however, is ensuring that KDE can be computed on-the-fly to support such flexible and adaptive interactions across scales---on which we focus in our work and for which we review past work next.

\subsection{Kernel density estimation (KDE)}
KDE \cite{diggle:1985:KDEDiggle,burt:2009:ESF} is a widely used method for computing data point distributions, which has been applied in diverse domains such as materials science~\cite{saito:2019:ASA}, astronomy~\cite{yuan:2020:AFM,yuan:2020:AFM:2}, geoscience~\cite{pelz:2023:ADB}, ecology~\cite{fleming:2017:ANK}, and traffic accident studies~\cite{xie:2008:KDET}.  Its core idea is that each data point is represented by a probability distribution (kernel; usually Gaussian~\cite{scott:2015:MDE} or Epanechnikov~\cite{epanechnikov:1969:Epanechnikov} ones) centered at its position, with the bandwidth controlling its spatial influence. Summing all kernels yields a continuous density field.

A key computational challenge in KDE lies in selecting the bandwidth, which controls the level of smoothing. The bandwidth can be specified as a constant (fixed KDE) or adapted to local point densities (adaptive KDE). Fixed KDE~\cite{brunsdon:1995:EPS,epanechnikov:1969:Epanechnikov} applies uniform smoothing across the dataset. While it is computationally less expensive, it often oversmooths dense regions and undersmooths sparse ones, limiting its accuracy in heterogeneous datasets. In contrast, adaptive KDE mitigates this issue by varying the bandwidth according to local density~\cite{brunsdon:1995:EPS, Silverman:1986:DEF}: points in sparse areas are assigned larger bandwidths, while those in dense regions receive smaller ones. This adaptive strategy yields more meaningful estimates in many applications, but comes at a high computational cost. For example, Brunsdon~\cite{brunsdon:1995:EPS} proposed a cross-validation-based method for determining spatially adaptive bandwidths, but its two-dimensional parameter search makes it inefficient for large datasets. 
\lixiang{For high-dimensional data, researchers proposed hashing-based estimators \cite{backurs:2019:SAT} to achieve sublinear query time and linear space and preprocessing time.}

\lixiang{KDE has been widely used in data visualization, with prior work exploring the algorithmic and system-level optimizations. Examples include GPU-accelerated KDE~\cite{lampe:2011:IVO} for interactive streaming visualization, efficient kernel density techniques~\cite{chan:2020:quad} for large-scale 2D hotspot visualization, adaptive GPU-based KDE methods~\cite{Zhang:2017:AGA} that combine algorithmic optimizations with parallel computation, and more recently, prefix-based spatiotemporal KDE frameworks~\cite{chan:2025:LSS} for large-scale, high-resolution density visualization. While these advanced KDE techniques offer improved theoretical efficiency and scalability and are tailored to different problem settings, our work focuses on immersive interaction integration and GPU-parallel implementation to support real-time, interactive multiscale exploration.}

\lixiang{KDE has also been leveraged to support in the context of immersive analytics. Prouzeau~\etal~\cite{prouzeau:2019:scaptics} introduced a VR-based technique that uses haptic feedback to convey KDE-derived density information, helping users identify occluded features in dense point cloud data.
Zhao~\etal~\cite{zhao:2023:metacast} employed KDE in immersive spatial selection techniques, enabling users to define selection regions based on density variations rather than individual points.}
These approaches, however, rely on precomputed single-scale density fields, limiting their ability to capture fine substructures across multiple levels of detail.
These efficiency limitations have motivated the exploration of high-performance computing strategies. 
An effective interaction with large-scale datasets, however, especially in immersive environments, requires not only scalability but also real-time, on-the-fly computation, which was not explicitly addressed in prior research.
To this end, we formulate a GPU-parallel algorithm based on a modified Breiman kernel density estimation method with a finite-support, adaptive Epanechnikov kernel, and further propose a parallel optimization method to accelerate adaptive bandwidth calculation, together improving the efficiency of the entire KDE pipeline.
To validate our approach, we demonstrate its use in two scientific data exploration scenarios and evaluate its effectiveness through both performance experiments and empirical user studies.

\section{\method: A dynamic KDE technique}
\label{sec:kde}

We begin our discussion with an overview of the KDE algorithm used in this work (\autoref{subsec:KDE}), then present \method, our GPU-accelerated KDE for multiscale analysis (\autoref{subsec:gpukde}), and finally report on experiments that evaluate its performance (\autoref{subsec:kde:performance}).

\subsection{Kernel density estimation}
\label{subsec:KDE}
\backuplineskips{}
\compactmathpar{%
KDE methods produce a smooth, continuous density field by distributing the contribution of each particle over a larger volume using a smoothing kernel function. 
\lixiang{This kernel function describes the spatial influence of each particle, assigning higher weights to locations closer to the particle center and lower weights to more distant locations. To estimate the scalar density field, similar to past works \cite{zhao:2023:metacast,Yu:2012:ESA,Yu:2016:CEE}, we apply the modified Breiman kernel density estimation method (MBE) with a finite-support adaptive Epanechnikov kernel~\cite{ferdosi:2011:CDE, wilkinson:1995:dataplot} as follows: 
The whole dataset or region of interest is enclosed within a bounding box $B$ and discretized into a uniform grid of resolution $\mathbf{res}$ (\eg, $128^3$), with grid nodes located at positions $\mathbf{r}^{(n)}$, where $n$ denotes the node index. 
For each spatial axis $k \in \{x,y,z\}$, we calculate the smoothing length $\ell$ as}}\extranegmathskip{}
\begin{equation}\label{eq:isl}
    \ell_{k} = \frac{2 \,\big(P_{k}^{(80)} - P_{k}^{(20)}\big)}{\log N}.
\end{equation}
\compactmathpar{\extranegmathskip{}%
Here, $N$ is the number of particles in the box $B$, and $P_{k}^{(q)}$ represents the $q$\textsuperscript{th} percentile of the coordinates along axis $k$. For the grid node at position $\mathbf{r}^{(n)}$, we calculate the pilot density $\rho_{\text{pilot}}(\mathbf{r}^{(n)})$ as:}\extranegmathskip{}
 \begin{equation}\label{eq:pilot}
     \rho_{\text{pilot}}(\mathbf{r}^{(n)})=\frac{15}{8 \pi N} \frac{1}{\ell_{x} \ell_{y} \ell_{z}} \sum_{j}  E(\|\mathbf{r}^{(n;j)}\|),
 \end{equation}
with
\begin{equation}
    \mathbf{r}_{k}^{(n ; j)}=(\mathbf{r}_{k}^{(n)}-\mathbf{r}_{k}^{(j)}) / \ell_{k},
\end{equation}
\compactmathpar{%
where $\mathbf{r}^{(j)}$ is the the $j$\textsuperscript{th} particle's position and $E(x)$ is the Epanechnikov kernel:}
 \begin{equation}\label{eq:e}
 E(x)=\left\{\begin{array}{ll}
 1-x^{2}, & |x| < 1, \\
 0, & |x| \ge 1. 
 \end{array}\right.
 \end{equation}
\compactmathpar{%
The function $E(x)$ implies that a particle contributes to pilot density $\rho_{\text{pilot}}(\mathbf{r}^{(n)})$ only if the particle $\mathbf{r}^{(j)}$ lies within the ellipsoid centered at the node position $\mathbf{r}^{(n)}$, with semi-axes $\ell_x$, $\ell_y$, and $\ell_z$.
\lixiang{This ellipsoidal restriction effectively limits the kernel bandwidth, ensuring that the resulting density estimate captures the local particle distribution.}
Then, we compute the pilot density $\rho_{\text{pilot}}(\mathbf{r}^{(j)})$ at the position of the $j$\textsuperscript{th} particle using multi-linear interpolation with respect to the densities on nearby nodes.
Next, we compute the mean pilot density $aveDen$ of all particles:}\extranegmathskip{}
\begin{equation}\label{eq:ave}
aveDen=\sum_{j}\rho_{\text{pilot}}(\mathbf{r}^{(j)})
\end{equation}
\compactmathpar{\extranegmathskip{}%
We then update the $j$-th particle’s smoothing length by}
\begin{equation} \label{eq:asl}
        \ell_k^{(j)} = \min\!\left(
        \ell_k \cdot \left(\frac{aveDen}{\rho_{\text{pilot}}(\mathbf{r}^{j})}\right)^{\frac{1}{3}},
        \; 5 \cdot \text{s}_k
    \right),
\end{equation}
\compactmathpar{%
where $\ell_k$ is the initial smoothing length along axis $k$, $s_k$ denotes the distance between adjacent nodes along the $k$-axis.
\lixiang{To prevent the kernel from becoming excessively large---which could increase computational cost and overly smooth the density---we apply an upper bound of $5 \cdot s_k$.}
Then, we re-evaluate the density on all the nodes based on the updated smoothing length.
For the grid node at position $\mathbf{r}^{(n)}$, we calculate the final density $\rho^(\mathbf{r}^{(n)})$ as:}
 \begin{equation}\label{eq:final}
     \rho(\mathbf{r}^{(n)})=\frac{15}{8 \pi N}  \sum_{j} \frac{1}{\ell_{x}^{(j)} \ell_{y}^{(j)} \ell_{z}^{(j)}} E(\|\mathbf{r}^{(n;j)}\|),
 \end{equation}
\compactmathpar{\extranegmathskip{}%
with}\extranegmathskip{}
\begin{equation}
    \mathbf{r}_{k}^{(n ; j)}=(\mathbf{r}_{k}^{(n)}-\mathbf{r}_{k}^{(j)}) / \ell_{k}^{(j)},
\end{equation}
Finally, we calculate the density $\rho(\mathbf{r})$ at any position $\mathbf{r}$ via trilinear interpolation of the surrounding grid nodes.
\restorelineskips{}

\subsection{\method: Fast and scalable KDE on GPU}
\label{subsec:gpukde}

To enable on-the-fly density evaluation, we implement the KDE pipeline on the GPU with key optimizations (\autoref{fig:app:flowchart} in \autoref{appendix-flowchart}). \lixiang{Our GPU parallel implementation follows HLSL conventions, with background on the GPU programming model provided in \autoref{appendix-gpuframe}.}
Prior to GPU execution, we perform essential preprocessing on the CPU: we determine the smoothing length (\autoref{eq:isl}) and create a k-d tree spatial index for the point cloud dataset to accelerate future neighborhood queries. We then upload the input data (k-d tree, smoothing length, particle positions, grid node positions, grid resolution, and node spacing) to GPU global memory.
In the GPU computation pipeline, as illustrated in HLSL-style pseudo-code in \autoref{alg:FSKDE}, we dispatch three compute kernels, each handling a distinct density estimation stage:
\begin{itemize}[nosep]
\item In \textbf{pilot density estimation} (\autoref{alg:pilot}, \autoref{subsec:gpukde:pilot}) we compute a pilot density field at grid nodes by a uniform smoothing length.
\item In \textbf{adaptive smoothing length} (\autoref{alg:asl}, \autoref{subsec:gpukde:asl}) we update per-particle smoothing lengths based on average pilot density.
\item In \textbf{final density estimation} (\autoref{alg:final}, \autoref{subsec:gpukde:final}) we recompute the density field using the adaptive smoothing lengths.
\end{itemize}

\begin{algorithm}[t]
\small
\caption{Fast and Scalable GPU-accelerated KDE}
\label{alg:FSKDE}
\Fn{\scriptsize\FSKDE{$kdTree, particles, nodes, res, l,s, density$}}{
\KwIn{$kdTree$---k-d tree spatial indexing structure}
\KwIn{$particles$---array of positions of all particles}
\KwIn{$nodes$---array of positions of all grid-nodes}
\KwIn{$res$---resolution of the grid (e.g. $64\times64\times64$ )}
\KwIn{$l$---array of smoothing length for every particle}
\KwIn{$s$---node spacing along x, y, z axis}
\KwOut{$density$---array of pilot density at grid-nodes}
\small$Dispatch\ (\ \textbf{PDE}\ (kdTree, particles, nodes, res, l, density), \frac{res_x}{PDE_{tx}}, \frac{res_y}{PDE_{ty}}, \frac{res_z}{PDE_{tz}});$

\small$Dispatch\ (\ \textbf{ASL}\ (particles, l,s, density), \frac{particles.\texttt{count}}{ASL_{tx}}, 1, 1);$\;
\small$Dispatch\ (\ \textbf{FDE}\ (particles, nodes, res, l, density), \frac{res_x}{FDE_{tx}}, \frac{res_y}{FDE_{ty}}, \frac{res_z}{FSDEtz});$\;
}
\vspace{-.5em}
\end{algorithm}

\subsubsection{Pilot density estimation (PDE)}
\label{subsec:gpukde:pilot}
To start, we compute the pilot density at each grid node by accumulating the contributions from neighboring particles.
We adopt the $\mathit{gather}$ approach~\cite{he:2007:EGA}, where each thread processes a grid node and collects contributions from neighboring particles, rather than a $\mathit{scatter}$ approach, where each thread processes a particle and distributes its contribution to surrounding nodes.
This way we avoid costly atomic operations and improve the performance and scalability for our large-scale particle datasets.

We dispatch the \textbf{PDE} kernel (\autoref{alg:pilot}) with \mathcompact{$\tfrac{\text{\textbf{res}}_{x}}{PDE_{tx}} \times \tfrac{\text{\textbf{res}}_{y}}{PDE_{ty}} \times \tfrac{\text{\textbf{res}}_{z}}{PDE_{tz}}$} thread groups, each consisting of $PDE_{tx} \times PDE_{ty} \times PDE_{tz}$ threads, to cover the entire grid domain. 
For each thread with global index $idx$, we evaluate the pilot density \mathcompact{$\rho_{\text{pilot}}(\mathbf{r}^{(idx)})$} at node position \mathcompact{$\mathbf{r}^{(idx)}$} using \autoref{eq:pilot}.
Since only particles within the kernel support contribute to the density (\autoref{eq:e}), we first identify candidate particles through a spherical range query on the k-d tree, centered at \mathcompact{$\mathbf{r}^{(idx)}$} with radius \mathcompact{$\max(\ell_x, \ell_y, \ell_z)$}, \ie, the largest of the three kernel semi-axes $\ell_x$, $\ell_y$, and $\ell_z$.
This query allows us to discard of particles outside the Epanechnikov kernel.
In this way, for $M$ nodes and $N$ particles, we reduce the complexity of the original KDE algorithm from $O(MN)$ to \mathcompact{$O(M\sqrt{N})$} (where k-d tree searching has complexity \mathcompact{$O(\sqrt{N}+K)$}, with $K$ the number of neighbors returned~\cite{kakde:2005:kdtree}).
Finally, we compute the pilot density \mathcompact{$\rho_{\text{pilot}}(\mathbf{r}^{(idx)})$} at node position \mathcompact{$\mathbf{r}^{(idx)}$} using \autoref{eq:pilot} and store it in the $idx$-th position of density buffer.

\begin{algorithm}[t]
\small
\caption{Pilot Density Estimation}
\label{alg:pilot}
$numthreads(PDE_{tx}, PDE_{ty}, PDE_{tz})$\;
\Fn{\scriptsize\PDE{$kdTree, particles, nodes, res, l, density$}}{
\KwIn{$kdTree,particles,nodes,res,l$}
\KwOut{$density$}
  $uint3 \ DTid : \text{SV\_DispatchThreadID};$ \;
  $uint \ idx = DTid_x + DTid_y \cdot res_x + DTid_z \cdot res_x \cdot res_y;$ \;
    \If{$idx \ge nodes.\texttt{count}$}{
    $return;$\;  }
  $float3 \ query = nodes[idx];$\;
  $float \ range = max(l[0]_x,l[0]_y,l[0]_z);$\;
  $uint \ retNbrs; $  \tcp{\scriptsize \textcolor{gray}{the number of returned neighbor particles}}
  $float \ retDists[]; $  \tcp{\scriptsize \textcolor{gray}{the array of neighbor–query distances}}
  $\mathrm{KDTreeSearch}(kdTree,\, query,\, range,\, retNbrs,\, retDists);$\;
  $float \ sum = 0;$\;
  \For{$i \ in \ 0:$  $retNbrs{-}1$}{
    $float3 \ r=float3(\frac{retDists[i]_x}{{l[0]}_x},\frac{retDists[i]_y}{{l[0]}_y},\frac{retDists[i]_z}{{l[0]}_z});$\;
    $sum+= \max(0, 1 - \|r\|^2)$\;
    $i++;$
  }
   $density[idx] =\frac{15 \cdot sum}{8\pi\cdot particles.count \cdot {l[0]}_x \cdot {l[0]}_y \cdot {l[0]}_z};$
}
\vspace{-.5em}
\end{algorithm}

\subsubsection{Adaptive smoothing length (ASL)}
\label{subsec:gpukde:asl}
Next, we update the smoothing length of each particle based on the result of pilot density estimation. 
The goal is to assign larger smoothing lengths in sparse regions and smaller ones in dense regions, thereby yielding a smoothly density field in the subsequent stage.
Since the update is particle-based, we assign one thread per particle to compute its adaptive smoothing length, which provides a natural and efficient mapping for parallel execution.

We dispatch the \textbf{ASL} kernel (\autoref{alg:asl}) with \mathcompact{$\tfrac{N}{ASL_{tx}} \times 1 \times 1$} thread groups, each containing \mathcompact{$ASL_{tx} \times 1 \times 1$} threads, to cover the entire grid domain.
For the thread with global index $idx$, we compute the adaptive smoothing length of the $idx$\textsuperscript{th} particle as follows.

\begin{algorithm}[!htbp]
\small
\caption{Adaptive Smooth Length}
\label{alg:asl}
$numthreads(ASL_{tx}, 1, 1)$\;
\Fn{\scriptsize\ASL{$particles, l,s, density$}}{
\KwIn{$particles,l,s,density$}
\KwOut{$l$}
  $uint \ GI : \text{SV\_GroupIndex};$ \;
  $uint3 \ GID : \text{SV\_GroupID}; $ \;
  $uint3 \ DTid : \text{SV\_DispatchThreadID};$ \;
  $uint \ idx = DTid_x; $ \;
  $float \ groupAveDen[\frac{particles.count}{ASL_{tx}}]; $ \;
  $groupshared \ float \ sharedDen[ASL_{tx}]; $ \;
    \If{$idx \ge particles.count$}{
        $return;$\;  }
  $float \ pDen=TrilinearInterp(particles[idx],density); $ \;
  $sharedDen[GI]=pDen; $ \;
  $GroupMemoryBarrierWithGroupSync();$ \;
  \tcp{\scriptsize \textcolor{gray}{parallel reduction}}
    \For{$i \ in \ ASL_{tx}/2:$ $0$}{            
        \If{$GI < i$}{
            $sharedDen[GI]+=sharedDen[GI+i];$\; 
            $GroupMemoryBarrier();$}
        $i>>=1;$\;
    } 
  \If{$GI == 0$}{
    $groupAveDen[GID_x]=sharedDen[0]/ASL_{tx};$}
    $WaitForAllThreads();$ \;
    $float \ aveDen = groupAveDen.Average();$  \;
    $l[idx]_x=\min\!\left(l[idx]_x \cdot \left|\frac{\textit{aveDen}}{\textit{pDen}}\right|^{\frac{1}{3}}, \; 5 \cdot \textit{s}_x \right);$ \;
    $l[idx]_y=\min\!\left(l[idx]_y \cdot \left|\frac{\textit{aveDen}}{\textit{pDen}}\right|^{\frac{1}{3}}, \; 5 \cdot \textit{s}_y \right);$ \;
    $l[idx]_z=\min\!\left(l[idx]_z \cdot \left|\frac{\textit{aveDen}}{\textit{pDen}}\right|^{\frac{1}{3}}, \; 5 \cdot \textit{s}_z \right);$ \;
}
\end{algorithm}

First, we compute the pilot density \mathcompact{$\rho_{\text{pilot}}(\mathbf{r}^{(idx)})$} of the $idx$\textsuperscript{th} particle using trilinear interpolation and stores the result as \texttt{pDen = TrilinearInterp(particles[idx], density)}.

Second, we calculate the arithmetic mean $aveDen$ of the pilot densities across every particle in box B, using \autoref{eq:ave} for efficient aggregation. 
To accelerate this step, we design a hierarchical computational method (HCM) that employs parallel reduction with shared memory, thereby reducing costly global memory accesses.
\hbox{Specifically}, within each thread group, we first allocate a shared memory array \texttt{sharedDen}, where each thread stores its locally computed pilot density. 
We then perform a parallel reduction within the group to  compute its average density, which is written to a global memory array \texttt{groupAveDen} with $N/ASL_{tx}$ entries. 
Then, we obtain the overall mean pilot density by averaging all values in \texttt{groupAveDen}.
Specifically, for thread with global index $idx$, along with other threads in the same thread group (group index $GID$), we proceed as follows:
\begin{enumerate}[nosep, leftmargin=2em]
    \item Thread $idx$ stores its pilot density $pDen$ into the shared memory array \texttt{sharedDen} at its local index $GI$ ($idx$\textsuperscript{th} thread's index within it's thread group), \ie, \texttt{sharedDen[GI] = pDen}.

    \item All threads in group $GID$ wait until every thread in the same group has updated \texttt{sharedDen} with its calculated pilot density, \ie, \texttt{GroupMemoryBarrierWithGroupSync()}.
    
    \item All threads in group $GID$ perform a parallel reduction on \texttt{sharedDen} to accumulate all density values. After \hbox{execution}, the total value is located at $0$\textsuperscript{th} position in \texttt{sharedDen} array.
    
    \item The $0$\textsuperscript{th} thread in group $GID$ computes the average value of \texttt{sharedDen} and then writes it to the global memory array \texttt{groupAveDen} at index $GID$, \ie, \texttt{groupAveDen[GID\_x] = sharedDen[0]/ASL\textsubscript{tx}}.

    \item All threads wait until every group's $0$\textsuperscript{th} thread has written its result to \texttt{groupAveDen}, \ie, \texttt{WaitForAllThreads()}.
    
    \item Thread $idx$ computes the global average pilot density $aveDen$ as the mean of all elements in \texttt{groupAveDen}, \ie, \texttt{aveDen = groupAveDen.Average()}.
\end{enumerate}

 Third, thread $idx$ computes the smoothing length of the $idx$\textsuperscript{th} particle using \autoref{eq:asl} and stores it in the smoothing length buffer.

\begin{algorithm}[!htbp]
\small
\caption{Final Density Estimation}
\label{alg:final}
$numthreads(FDE_{tx}, FDE_{ty}, FDE_{tz})$ \;
\Fn{\scriptsize\FDE{$particles, nodes, res, l, density$}}{
\KwIn{$particles,nodes,res,l$}
\KwOut{$density$}
  $uint3 \ DTid : \text{SV\_DispatchThreadID};$ \;
  $uint \ idx = DTid_x + DTid_y \cdot res_x + DTid_z \cdot res_x \cdot res_y;$ \;
    \If{$idx \ge nodes.count$}{
        $return;$\;  }  
  $float \ sum = 0;$\;
    \For{$i \ in \ 0:$ $particles.count{-}1$}{
        $float3 \ r;$\;
        $r_x=\frac{nodes[idx]_x-particles[i]_x}{l[i]_x};$\;
        $r_y=\frac{nodes[idx]_y-particles[i]_y}{l[i]_y};$\;
        $r_z=\frac{nodes[idx]_z-particles[i]_z}{l[i]_z};$\;
        $sum+= \frac{\max(0, 1 - \|r\|^2)}{l[i]_x \cdot l[i]_y \cdot l[i]_z} ;$\;
        $i++;$
        }
  $density[idx] =\frac{15 \cdot sum}{8\pi\cdot particles.count };$
}
\vspace{-.5em}
\end{algorithm}

\subsubsection{Final density estimation (FDE)}
\label{subsec:gpukde:final}
Finally, we recompute the density at each node using the adaptive smoothing length of its contributing particles.
We dispatch the \textbf{FDE} kernel (\autoref{alg:final}), where each thread computes the final density at a single grid node. 
To cover the grid, we launch $\tfrac{\mathbf{res}_x}{FDE_{tx}} \times \tfrac{\mathbf{res}_y}{FDE_{ty}} \times \tfrac{\mathbf{res}_z}{FDE_{tz}}$ thread groups, each containing $FDE_{tx} \times FDE_{ty} \times FDE_{tz}$ threads.
The thread with global index of $idx$ evaluates the final density $\rho(\mathbf{r}^{(idx)})$ at node $\mathbf{r}^{(idx)}$ using \autoref{eq:final} and stores the result to the density buffer at position $idx$.
After all threads have completed, we transfer the density buffer from GPU to CPU memory.

\subsection{Performance analysis}
\label{subsec:kde:performance}
Before we report our empirical results, we briefly analyze the algorithmic complexity of \method to highlight its fundamental sources of speedup. These optimizations target the density field computation, which requires recomputation whenever the view changes (\eg, transitioning to a new scale). For pilot density computation, we use a k-d tree to reduce neighborhood search complexity from $O(MN)$ for $M$ grid nodes and $N$ particles to \mathcompact{$O(M\sqrt{N})$}. To achieve an adaptive smoothing length, we employ a hierarchical method that utilizes thread-group shared memory and parallel reduction, thereby lowering the aggregation complexity from $O(N)$ to $O(\log N)$ within each group of size $N$. Combined with GPU parallelism, these optimizations enable \method to scale efficiently to hundreds of thousands of particles
and deliver low-latency density field recomputation.
To demonstrate this performance in practice, next we evaluate our approach on astronomical point cloud data.

\textbf{Design.}
To assess the \textit{efficiency} of \method, following prior work~\cite{Zhang:2017:AGA}, we implemented three density estimation strategies: a sequential version on a single CPU core (SC) as baseline, a parallel version on a multi-core CPU (MC), and the GPU-parallel version (\method) as described in \autoref{subsec:gpukde}.
\lingyun{These CPU-based baselines provide a controlled reference that allows us to establish real-time feasibility and to attribute performance differences specifically to GPU parallelization and dynamic recomputation, rather than to algorithmic variation.}
We compared execution time across the three strategies using the acceleration factor (AF)~\cite{zhang:2016:EPP} as a performance metric, defined as $AF = T_{baseline} / T_{target}$.
We report the average execution time over 10 runs for all three strategies, with GPU-parallel \method results excluding CPU–GPU data transfer.

\textbf{Environment and implementation.}
We conducted the experiment on a workstation running Windows~11 on Intel's 13\textsuperscript{th} generation Core\texttrademark{} i9-13900KF processor (3.0~GHz, 64~GB RAM) and an NVIDIA GeForce RTX~4090 GPU (24~GB of memory) for acceleration. 
We implemented \method in Unity3D, with GPU computations realized through compute shaders. We set the resolution $res$ of box $B$ to $64^3$ (262k nodes) and the configuration of the thread group dimensions (\ie, $PDE_{tx},PDE_{ty},PDE_{tz},FDE_{tx},FDE_{ty},FDE_{tz}$) as 8 and $ASL_{tx}$ as 1024.
In addition, we implemented the sequential CPU strategy (SC) using C$\#$ and the multi-core CPU strategy (MC) using C$\#$'s Task Parallel Library (TPL).

\begin{figure}[t]
    \centering
    \begin{subfigure}{0.3\linewidth}
        \begin{tikzpicture}
            \node[inner sep=0] (img)
                {\includegraphics[width=\linewidth]{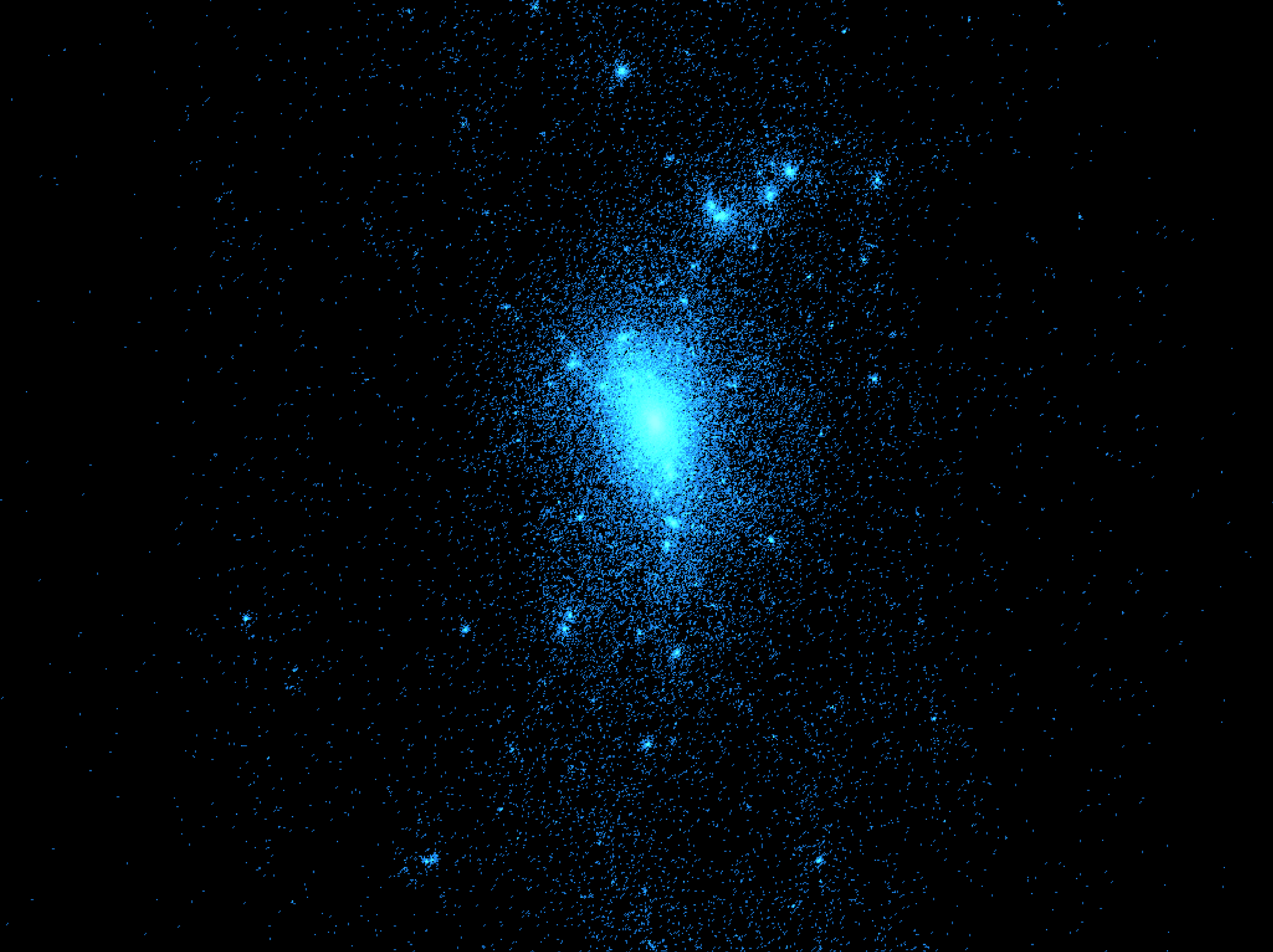}};
            \node[anchor=south west, text=white, font=\bfseries, inner sep=1pt,
                  fill=black, fill opacity=0.6, text opacity=1]
                at ([xshift=2pt,yshift=2pt]img.south west) {(a)};
        \end{tikzpicture}
        \caption{} \label{fig:nbody1}
    \end{subfigure}\hfill
    \begin{subfigure}{0.3\linewidth}
        \begin{tikzpicture}
            \node[inner sep=0] (img)
                {\includegraphics[width=\linewidth]{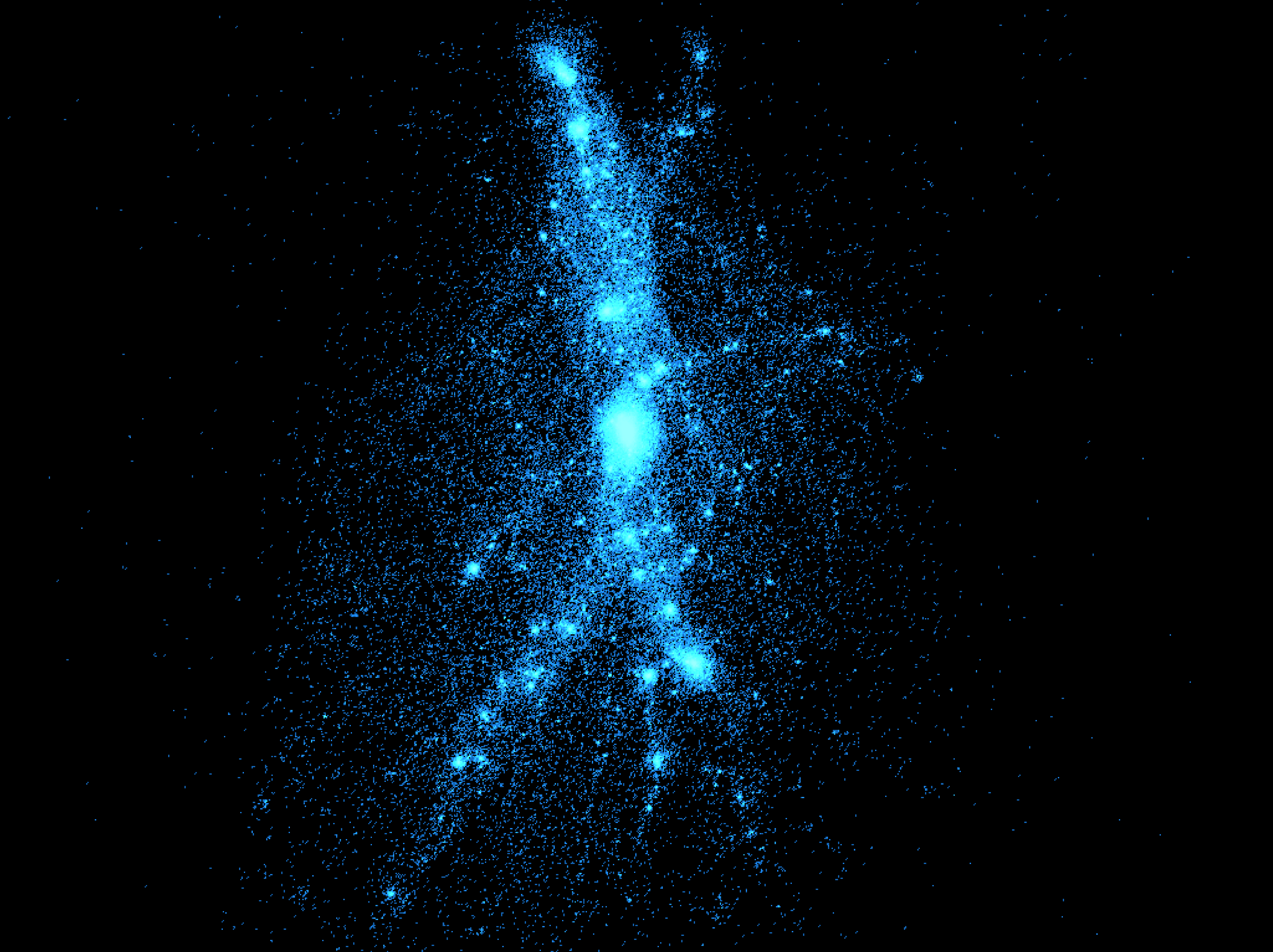}};
            \node[anchor=south west, text=white, font=\bfseries, inner sep=1pt,
                  fill=black, fill opacity=0.6, text opacity=1]
                at ([xshift=2pt,yshift=2pt]img.south west) {(b)};
        \end{tikzpicture}
        \caption{} \label{fig:nbody2}
    \end{subfigure}\hfill
    \begin{subfigure}{0.3\linewidth}
        \begin{tikzpicture}
            \node[inner sep=0] (img)
                {\includegraphics[width=\linewidth]{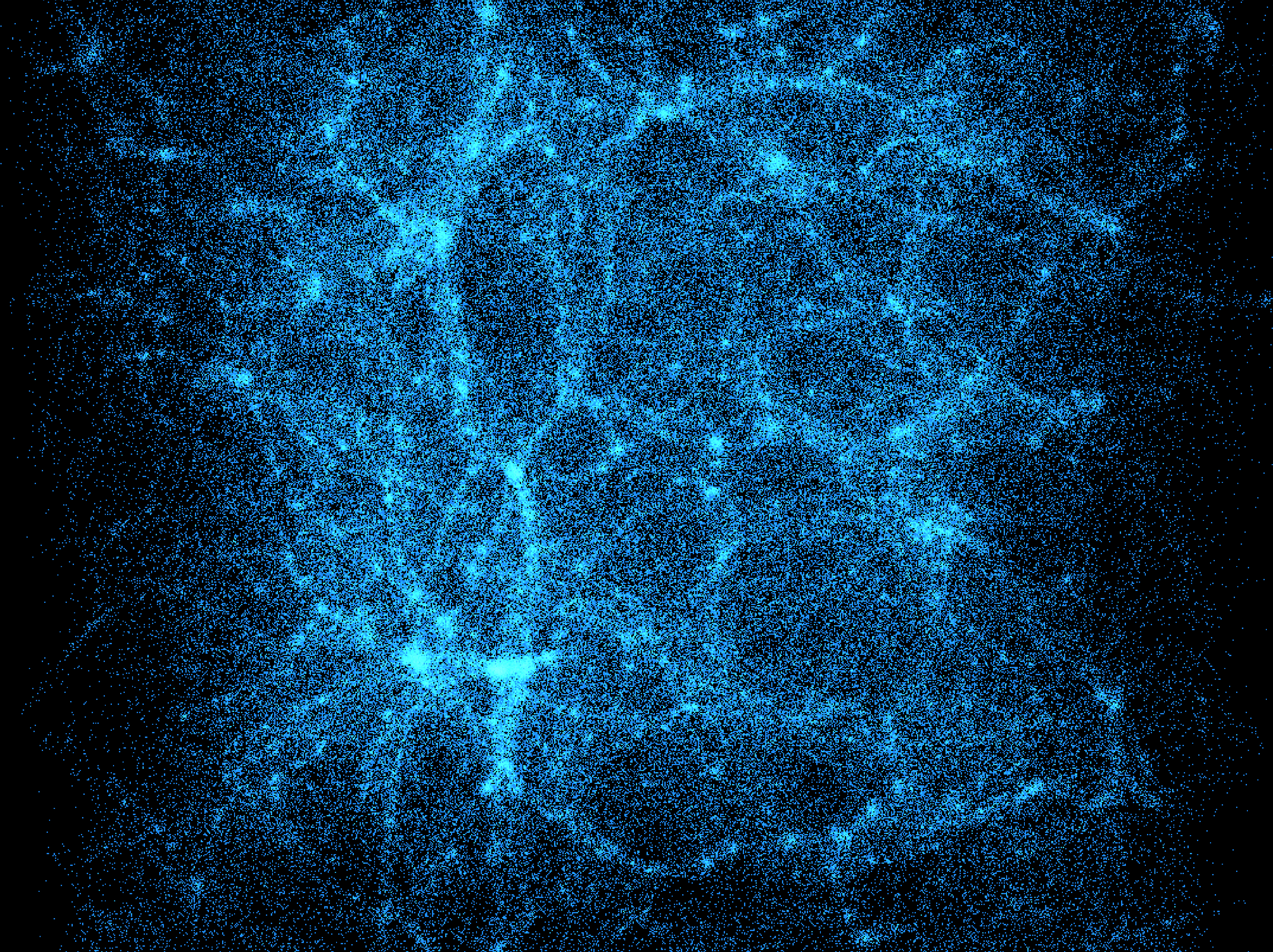}};
            \node[anchor=south west, text=white, font=\bfseries, inner sep=1pt,
                  fill=black, fill opacity=0.6, text opacity=1]
                at ([xshift=2pt,yshift=2pt]img.south west) {(c)};
        \end{tikzpicture}
        \caption{} \label{fig:filament}
    \end{subfigure}
    \vspace{-3ex}
    \caption{Experiment datasets: (a) Nbody 1 (76k points), 
             (b) Nbody 2 (164k points), and 
             (c) Filament (442k points).}
    \label{fig:exp_data}
\end{figure}

\textbf{Datasets.}
We used three cosmological point cloud datasets: Nbody 1 (76k points, \FigSubref{fig:exp_data}{fig:nbody1}), Nbody 2 (164k points, \FigSubref{fig:exp_data}{fig:nbody2}), and Filament (442k points, \FigSubref{fig:exp_data}{fig:filament}). The two N-body datasets feature a dense central cluster surrounded by smaller ones \cite{springel:2008:aquarius}, while the filament dataset depicts a cosmic web with thin filaments\cite{Springel:2005:SimulationsOT}.

\textbf{Efficiency.}
Our results (\autoref{tab:exp_effi}) show that \method ran substantially
faster than SC and MC, with standard deviation consistently within 5\% of the mean, indicating stable performance. 
As dataset size increases, \method becomes increasingly more efficient relative to CPU implementations, with the acceleration factor rising accordingly.
For the 76K-point dataset discretized on a $64^3$ grid, \method achieved an execution time of 0.042~s, corresponding to a sustained rate of about 20~FPS if executed continuously. 
Although the computation time does not meet the standard 60 FPS, density recomputation only needs to be triggered occasionally, when the view changes substantially. Moreover, the execution times of \method were generally in the order of 0.3--0.1s or substantially lower, so that the ``illusion of animation'' remains unaffected \cite{Card:1991:TIV}.
The overall runtime and system fluidity thus remain unaffected.

\newlength{\columnspacing}
\newlength{\numberwidth}
\setlength{\columnspacing}{4pt}
\begin{table}[t]
\small
\settowidth{\numberwidth}{0}
\centering
\caption{Execution time (seconds) and acceleration factor of SC, MC (32 cores), and \method on a $64^3$ grid (262k nodes).}\vspace{-.5\baselineskip}
\label{tab:exp_effi}
\setlength{\tabcolsep}{10pt}
\begin{tabu}{@{}l@{\hspace{\columnspacing}}l@{\hspace{\columnspacing}}c@{\hspace{\columnspacing}}c@{\hspace{\columnspacing}}c@{\hspace{\columnspacing}}c@{\hspace{\columnspacing}}c@{\hspace{\columnspacing}}c@{}}
\toprule
\multirowcell{3}[-0.8ex]{metric} & \multirowcell{3}[-0.8ex]{algorithm} & 
\multicolumn{2}{@{\hspace{\columnspacing}}c@{\hspace{\columnspacing}}}{Nbody1 (76k)} & \multicolumn{2}{@{\hspace{\columnspacing}}c@{\hspace{\columnspacing}}}{Nbody2 (164k)} & \multicolumn{2}{@{\hspace{\columnspacing}}c@{\hspace{\columnspacing}}}{filament (442k)} \\
\cmidrule(lr){3-4} \cmidrule(lr){5-6} \cmidrule(lr){7-8}
 & & mean & std.\ dev. & mean & std.\ dev. & mean & std.\ dev. \\
\midrule
\multirow{3}{*}{\begin{minipage}{14mm}execution\\ time (s)\end{minipage}} 
& SC      & 7.689 & 0.161 & 58.373 & 6.643 & 395.675 & 12.458 \\
& MC      & 1.542  & 0.041  & \hspace{\numberwidth}7.244  & 0.264  & \hspace{\numberwidth}48.514   & \hspace{\numberwidth}1.567  \\
& \method & 0.042  & 0.002 & \hspace{\numberwidth}0.119  & 0.003 & \hspace{2\numberwidth}0.309   & \hspace{\numberwidth}0.011 \\
\midrule
\multirow{2}{*}{\begin{minipage}{14mm}acceleration\\ factor\end{minipage}} 
& SC/\method & 183.1 & -- & 409.5 & -- & 1280.5 & -- \\
& MC/\method & \hspace{\numberwidth}36.7  & -- & \hspace{\numberwidth}60.8  & -- & \hspace{\numberwidth}157.0  & -- \\
\bottomrule
\end{tabu}
\end{table}

\section{Multiscale exploration with \method}
\label{sec:exploration}

We now discuss how our dynamic KDE approach contributes to improving large-scale scientific data exploration in VR. 

\subsection{Scalable selection technique}
\label{subsec:selection:method}
Selection is a fundamental operation for many subsequent data analyses~\cite{wills:1996:S5W,besanccon:2021:state-of-art}. Maintaining both efficiency and accuracy, however, is particularly challenging in multiscale point cloud datasets due to their inherent features: highly complex structures, multiple levels of scale, occlusion, and unclear or overlapping boundaries. 
Designing selection techniques for multiscale point cloud data requires an understanding of user interaction. Users navigate across scales to find a suitable view that allows them to see targets clearly. For ease of use, however, they may also select point clouds even when not at the ideal scale. Selection must thus be supported anywhere and at any scale. As users move across scales, furthermore, they may not want to manually adjust selection parameters---they expect these adjustments to happen automatically. In sum, a selection technique should be dynamic, adaptive, accurate, and efficient, which is exactly what we can realize by fitting our dynamic KDE into the selection workflow. Workflow visualization available in supplementary materials.

At the start of an interactive session, the global view is initialized. As users navigate to a new scale, we recompute the density field on the GPU using our KDE~(\autoref{subsec:gpukde}) and then transfer it back to the CPU for subsequent operations. Although we update the density field whenever the scale changes, the GPU pipeline enables rapid recomputation, ensuring smooth navigation.

When users initiate a selection, we invoke a GPU kernel to compute the selection volume. Specifically, once a selection is initialized, we determine a density threshold for the interacting area and transfer it to the GPU. A kernel then executes the Marching Cubes algorithm~\cite{Wyvill:1986:DSS,Lorensen:1987:MCH} to extract the selection volume as an iso-surface based on this threshold. We then select points within the extracted volume. Thanks to our optimized computation, the selection volume is generated rapidly. Even with the MeTAPoint technique~\cite{zhao:2023:metacast}, which detects density thresholds on-the-fly at the user's interaction point, users can immediately observe updates to the selection volume and assess whether the result aligns with their intent.

Overall, \lingyun{our framework supports adaptive and smooth scale transitions during selection by recomputing density fields on the fly to match users' interactive focus. This enables users to navigate and zoom across scales without noticeable delay.} Since the density field is recomputed with our optimized KDE approach once after a view change and remains available for subsequent selections, the technique supports fluid interaction at any scale. 

\subsection{Progressive navigation technique}
\label{subsec:navigation:method}
\begin{figure}[t]
    \centering
    \includegraphics[width=.8\linewidth]{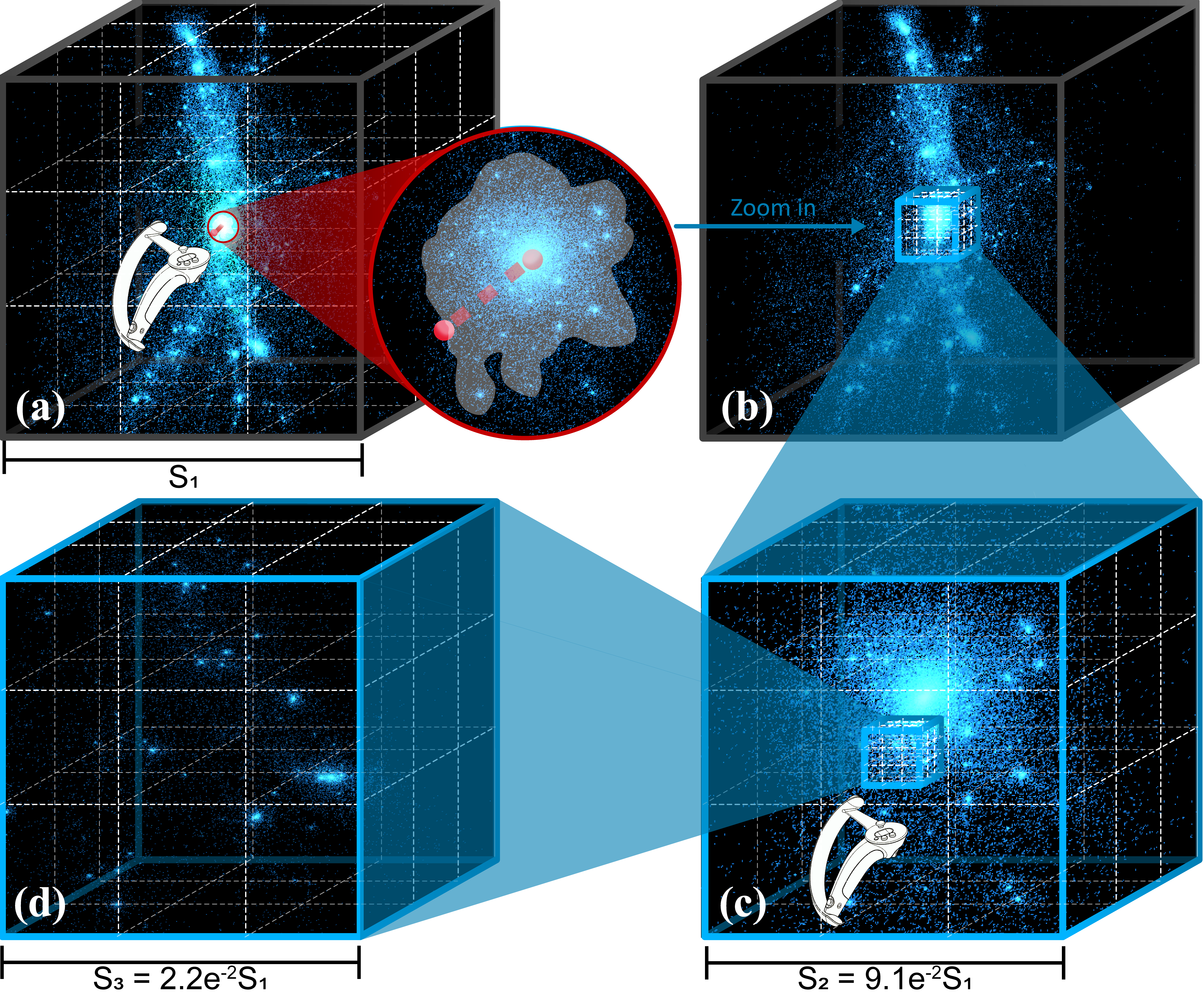}
    \caption{Our progressive navigation technique.}
    \label{fig:navigation}
    \vspace{-1em}
\end{figure}

Beyond being a selection technique, in fact, \method also integrates navigation and zooming into a unified workflow. 
Navigation is a fundamental exploration task that relocates the viewpoint toward a region of interest while preserving spatial context. It has been widely studied in visualization research across different scenarios, such as gaining insights from volumetric~\cite{Hsu:2013:AMC} and abstract data~\cite{Yang:2021:ENI}. Navigation techniques consist of several components, including wayfinding, travel, and context switching. In large multiscale point clouds, navigation becomes particularly challenging because salient structures are distributed across multiple scales, and dense spatial distribution introduces severe occlusion and clutter. These issues complicate wayfinding and travel, increase the number of travel steps that often occur across scales, and intensify context switching, making it easy for users to lose orientation and spatial context when moving across scales. Navigating such data in immersive environments is even more demanding, as the strong sense of presence can amplify both perceptual challenges and spatial disorientation. To address these challenges, we propose a scale-aware progressive navigation technique that enables users to explore multiscale point cloud data across scales. By defining a region of interest (ROI)---essentially a selection interaction that we have just described---our technique guides users through successive scales (\ie, a combined navigation and zooming interaction) to reach the most suitable viewpoint for observation. Next, we demonstrate how our dynamic KDE can fit into the navigation/zooming workflow to support this process. A workflow visualization is provided in \autoref{appendix-flowchart}.

An interactive session starts from a global view of the dataset. As before, we recompute the density field on the GPU using our KDE (\autoref{subsec:gpukde}) whenever the view changes substantially. To navigate/zoom to a specific area, users define an ROI with our scalable selection technique (\autoref{fig:navigation}(a), \autoref{subsec:selection:method}). Because we compute the density field at the current scale, the ROI is immediately identified and we can smoothly transition the viewpoint to the ROI's center (from \autoref{fig:navigation}(b) to (c)). 
We adjust the size of the new viewport to adequately cover the ROI (\eg, the identified selection volume), guiding users to an appropriate scale for observation. \lingyun{Together, this process supports stable viewpoint-driven exploration by preserving spatial orientation and contextual continuity across scale changes.}
While the current implementation focuses on the orientation consistency, a promising extension could also be to combine our scale-aware navigation with feature-aware techniques (\eg, \cite{Cao:2025:FBV,Yang:2019:DLB}) to automatically compute an optimal view orientation, enabling users to travel across scales and arrive at the most informative perspective.

After each viewpoint update, we recompute the density field for the new scale, allowing users to effectively perform further ROI selections that guide subsequent navigation steps (as shown in \autoref{fig:navigation}(d)). This progressive ``\emph{targeting}-\emph{travel}-\emph{context switching}'' workflow ensures smooth viewpoint transitions and multiscale navigation without the need for manual wayfinding. In addition, these travel steps can be recorded to enhance spatial awareness during progressive navigation. Overall, our dynamic KDE enables seamless, scale-aware navigation by continuously updating density fields to support progressive ROI selection and smooth view transitions.

\section{User study}
\label{sec:userstudy}

To evaluate whether our scalable selection method meets the goals of being dynamic, adaptive, accurate, and efficient, we conducted a within-subjects controlled user study comparing it against two density-based strategies: one using a precomputed single resolution and the other using precomputed multiple resolutions. 
We focus on the selection interaction, as it serves as the foundation for both pure data selection (\autoref{subsec:selection:method}) and selection-based progressive data navigation/zooming (\autoref{subsec:navigation:method}). We pre-registered our study (\href{https://osf.io/hfu6e}{\texttt{osf\discretionary{}{.}{.}io\discretionary{/}{}{/}hfu6e}}) and \lixiang{received IRB approval for the protocol (XJTLU University Research Ethics Review Panel, \textnumero~ER-LRR-0010000120520250813002121).}

\subsection{Study Design}
\label{subsec:stu_de}
\textbf{\textit{Participants.}}
We recruited $24$ unpaid participants ($11$ male, $13$ female) from the local university, aged $19$--$30$ years (M=$23.25$, SD=$3.22$). Among them, 21 were right-handed, two were left-handed, and one was ambidextrous.
Twelve participants reported using VR at least once per week, 11 at least once per year, and one had no prior experience with VR devices. Furthermore, 16 participants had obtained a Bachelor's degree or higher. All participants had normal \lixiang{($n=16$)} or corrected-to-normal \lixiang{($n=8$)} vision and could clearly distinguish the colors we used in our study.

\textbf{\textit{Apparatus.}}
We used the Vive Pro 2~\cite{htcvivepro2}, a PC-based VR head-mounted display (HMD; 2448\,\texttimes\,2448 resolution per eye, 116\textdegree{} field of view, 120\,Hz refresh rate). The study was carried out on a PC (Intel 13th Gen Core\texttrademark{} i9-13900KF processor, 3.0\,GHz, 64\,GB RAM and an NVIDIA GeForce RTX~4090 GPU, 24\,GB of memory). 

\textbf{\textit{Datasets.}}
We extracted five timesteps from a cosmological N-body simulation~\cite{springel:2008:aquarius} and used them as our datasets. These datasets feature stellar clusters distributed across multiple scales, where zoomed-in views reveal progressively detailed substructures. Each dataset contained eight to ten target structures highlighted in \textcolor{Goldenrod}{yellow}, distributed across different scales. Tasks began at varying scales, with all targets becoming visible as participants zoomed in. We show an example dataset with its target structures in \autoref{fig:UserStudyDatasets}, and other four datasets in the \autoref{appendix-dataset_userstudy}.

\begin{figure}
    \centering
    \includegraphics[width=\linewidth]{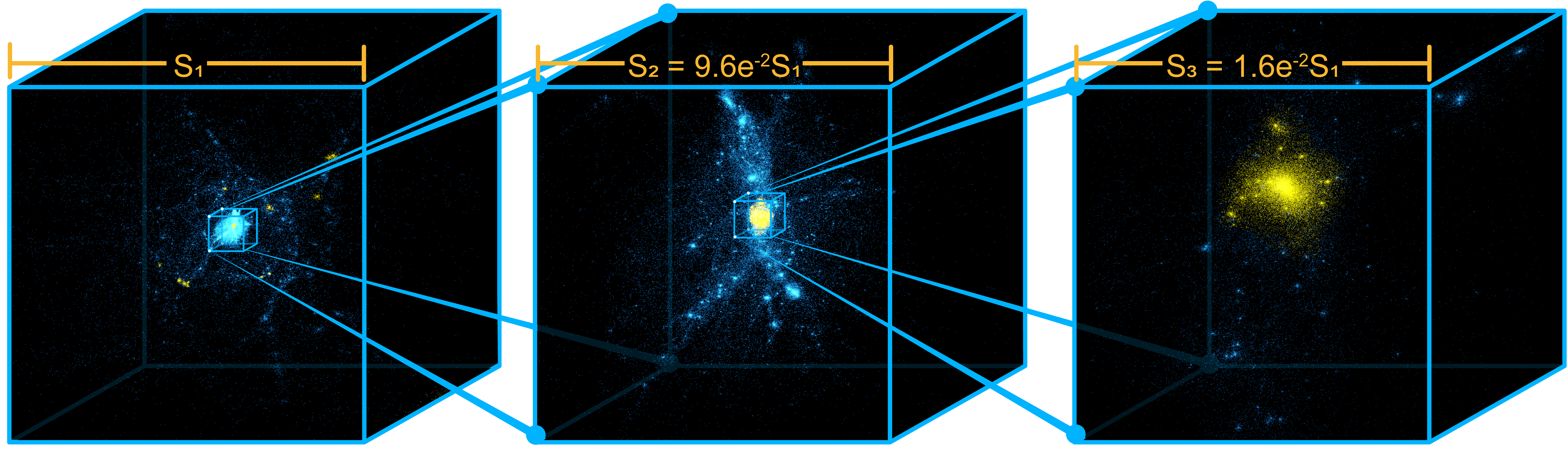}
    \caption{One of the datasets we used in our study.}
    \label{fig:UserStudyDatasets}
    \vspace{-1em}
\end{figure}

\textbf{\textit{Task and Procedure.}}
We instructed participants to select the \textcolor{Goldenrod}{yellow} particles while avoiding the \textcolor{DodgerBlue}{blue} ones. 
Prior to the main experiment and after we had obtained their informed consent, we trained them on the MeTAPoint selection technique \cite{zhao:2023:metacast} using practice datasets. Unlike what Zhao et al.\ \cite{zhao:2023:metacast} did in their MeTAPoint study, we allowed participants to freely adjust scales (via zooming) and to perform selections at different levels. We also provided undo, redo, and reset functions to restore the initial state if needed. 
In the main experiment, we asked participants to complete the selections as quickly and accurately as possible. We did not, however, provide suggestions on whether or when the selection was complete.
After each condition, we asked participants to report their workload and fatigue using NASA’s Task Load Index (TLX)~\cite{TLXScale}. At the end of all trials, we asked them to indicate their preferred technique and explain their choice, focusing on perceived fluency in multiscale selections. \lingyun{The interview questions are provided in \autoref{appendix-inter_que}. All participant responses are shared on OSF.} A whole session lasted approx.\ $45$ minutes. 

\textbf{\textit{Density Field Conditions.}}
We asked our participants to perform all selections using the same den\-si\-ty-based MeTAPoint method~\cite{zhao:2023:metacast}: they initiated a selection by pointing at or near a target cluster and then dragged along its boundary while holding the VR controller trigger. The underlying density field computation, however, differed across approaches, resulting in three experimental conditions:
\begin{description}[nosep]
\item[Precomputed Single-resolution (PS):] The density field is precomputed once at a fixed resolution ($64^3$) before the task. 
During the task, participants interact with this static field, which remains unchanged across all scale levels.
\item[Precomputed Multi-resolution (PM):] The density field is precomputed at two resolutions ($64^3$ and $128^3$). 
Users interact with the precomputed field interpolated by fields according to their scale level. 
We use two reference scales, $S_{max}$ and $S_{min}$, which are predefined---at $S_{min}$ the highest-resolution field is used, whereas at $S_{max}$ the lowest-resolution field is employed. 
For the current scale $S_c$ during exploration we determine the corresponding mipmap level $ml \in [0,1]$ by\\[0pt]
\begin{minipage}{\linewidth}
\begin{equation}
ml = \frac{S_c - S_{min}}{S_{max} - S_{min}} .
\end{equation}
\end{minipage}\\[3pt]
and use $ml$ to obtain the immediate density field via linear interpolation between the two fields with different resolutions.
\item[Dynamic Resolution (DR):] We compute the density field in real time via \method (\autoref{subsec:selection:method}). 
We set the field resolution to $64^3$ and configure the thread group dimensions as $PDE_{tx}$, $PDE_{ty}$, $PDE_{tz}$, $FDE_{tx}$, $FDE_{ty}$, $FDE_{tz}=8$, and $ASL_{tx}=1024$.
\end{description}
\lixiang{While PS represents the conventional approach of using a fixed-resolution density field for density-based selection~\cite{Yu:2012:ESA,Yu:2016:CEE,zhao:2023:metacast}, PM further optimizes this strategy by adopting a mipmap-inspired hierarchy from computer graphics, enabling multiscale access through precomputed density levels without real-time recomputation. By comparing DR against these two representative paradigms, we assess the performance of dynamic KDE in selection tasks.}

\textbf{\textit{Design.}} We counterbalanced the three density field conditions using full permutation (in total six possible orders). We assigned the first six participants one order each based on $P_{ID} \bmod 6$, and repeated the scheme for every subsequent group of six. In total, 24 participants\,\texttimes\,3 methods\,\texttimes\,5 datasets yielded 360 trials.

\textbf{\textit{Measures and Analysis.}} We recorded accuracy, completion time, as well as the transition times in the study. Given the criticism of NHST in the analysis of experimental data \cite{Baguley:2009:ES, Cumming:2013:NS, Dragicevic:2016:FSC, Dragicevic:2014:RAH}, and APA's advice to seek other methods \cite{VandenBos:2009:PMAPA}, we present our findings using estimation techniques that report effect sizes and confidence intervals instead of relying on $p$-value statistics.

\begin{description}[nosep]
\item[Accuracy:] Similar to Yu \etal~\cite{Yu:2012:ESA,Yu:2016:CEE,zhao:2023:metacast}, we evaluated the accuracy using two metrics: F1 and MCC (Matthews correlation coefficient). To calculate these, we identified true positives (TP; correctly selected particles), false positives (FP; incorrectly selected particles), false negatives (FN; target particles that were not selected), and true negatives (TN; correctly unselected particles). We defined precision as $P = TP/(TP+FP)$ and recall as $R = TP/(TP+FN)$, and calculated F1 as ${\mathrm{F1}=2\cdot (P \cdot R)/(P+R)}$. While F1 reflects the harmonic mean of precision and recall, it does not account for TN. Thus, we also computed MCC, defined as:\\[-5pt]
\begin{minipage}{\linewidth}
\begin{equation}
\mathrm{MCC} = \frac{TP \cdot TN - FP \cdot FN}{\sqrt{(TP+FP)(TP+FN)(TN+FP)(TN+FN)}}.
\notag
\end{equation}
\end{minipage}\\[3pt]
We normalized all accuracy scores, and computed means and 95\% bootstrap confidence intervals (CIs; $n$ $=$ 24).
\item[Completion Time:]
 We analyzed completion times using exact CIs on log-transformed data ($n$ $=$ 24). We report results as geometric means and demonstrate means comparisons as ratios.
\end{description}

\subsection{Hypotheses}
\label{sub:userstudy:hypo}
We formulated the following hypotheses based on the underlying principles of each density field condition:
\label{subsec:Hypotheses}
\newlength{\hypothesislabelwidth}%
\settowidth{\hypothesislabelwidth}{\textbf{H1}}%
\addtolength{\hypothesislabelwidth}{1.5ex}
\begin{description}[nosep]
\item[H1:] DR yields higher accuracy than PM and PS.  
\item[H2:] PM yields higher accuracy than PS.
\item[H3:] DR requires less completion time than PM and PS.
\item[H4:] PM requires less completion time than PS.
\item[H5:] DR has a lower cognitive load than PM and PS.
\item[H6:] DR has a higher preference than PM and PS.
\end{description}

The rationale behind \textbf{H1} is that density fields in DR are dynamically adjusted to the current scale during navigation and zooming. At any scale, the density field is recomputed with a resolution $64^3$, ensuring consistent accuracy. By contrast, PS and PM rely on pre-computed density fields: when zoomed in, they use coarser grids that reduce precision. 
Nevertheless, PM integrates density fields at two scales rather than just one as in PS. Therefore, selections made with PM should be more accurate than those with PS, as stated in \textbf{H2}.
The rationale behind \textbf{H3} is that DR provides the most accurate results, making selections easier for users to accept and thereby reducing completion time. Although the density fields in PS and PM are pre-computed and do not require on-the-fly computation, users may spend additional time refining their selections to achieve greater precision.
A similar reasoning applies to \textbf{H4}: because PM yields more accurate results than PS, users are more likely to accept the selection outcomes with less refinement, resulting in shorter completion times.
In addition, DR enables users to select target points accurately at any scale without the need to find an optimal view or repeatedly refine their selections, which reduces both effort and mental demand---thereby lowering cognitive load (\textbf{H5}). Consequently, users are more likely to prefer DR over PS and PM (\textbf{H6}).

\begin{figure}[t]
    \centering

    \begin{subfigure}{\columnwidth}
        \begin{tikzpicture}
            \node[inner sep=0] (img){\includegraphics[width=\columnwidth]{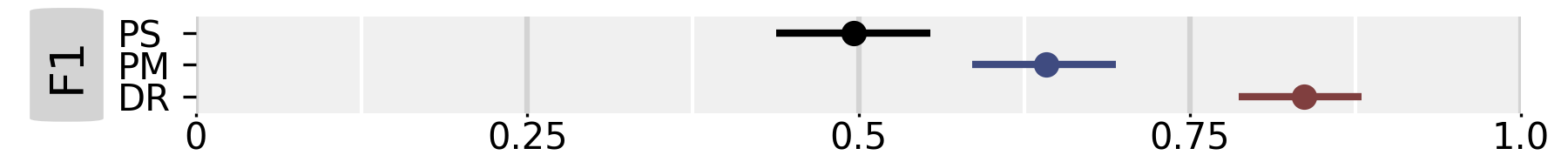}};
            \node[anchor=south west, inner sep=1pt, fill=white, fill opacity=.8, text opacity=1]
                at ([xshift=-8pt,yshift=3pt]img.south west) {\footnotesize (a)};
        \end{tikzpicture}
        \caption{}\label{fig:F1_Mean} 
    \end{subfigure}
    
    \vspace{-5.5ex}

    \begin{subfigure}{\columnwidth}
        \begin{tikzpicture}
            \node[inner sep=0] (img){\includegraphics[width=\columnwidth]{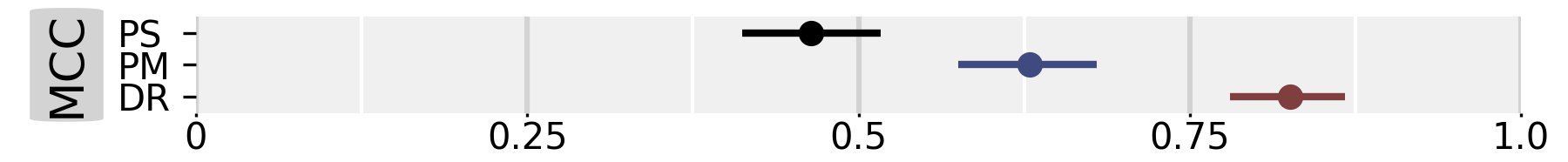}};
            \node[anchor=south west, inner sep=1pt, fill=white, fill opacity=.8, text opacity=1]
                at ([xshift=-8pt,yshift=3pt]img.south west) {\footnotesize (b)};
        \end{tikzpicture}
        \caption{}\label{fig:MCC_Mean}
    \end{subfigure}

    \vspace{-4.4ex}

    \begin{subfigure}{\columnwidth}
        \begin{tikzpicture}
            \node[inner sep=0] (img){\includegraphics[width=\columnwidth]{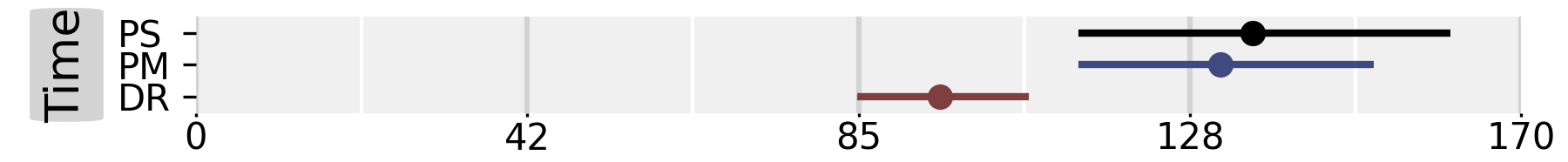}};
            \node[anchor=south west, inner sep=1pt, fill=white, fill opacity=.8, text opacity=1]
                at ([xshift=-8pt,yshift=3pt]img.south west) {\footnotesize (c)};
        \end{tikzpicture}
        \caption{}\label{fig:Time_Mean}
    \end{subfigure}
    \vspace{-6.5ex}
    \caption{The mean of (a) F1, (b) MCC, and (c) time, with 95\% CIs.}
    \label{fig:Mean}
\end{figure}

\vspace{-2ex}

\subsection{Results}
\label{subsec:Results}
\begin{figure}[t]
    \centering

    \begin{subfigure}{\columnwidth}
        \begin{tikzpicture}
            \node[inner sep=0] (img){\includegraphics[width=\columnwidth]{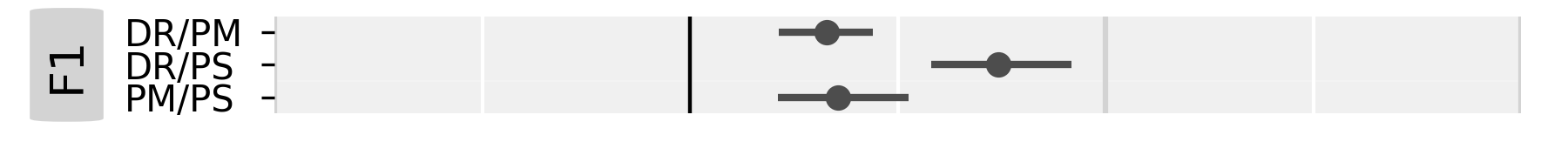}};
            \node[anchor=south west, inner sep=1pt, fill=white, fill opacity=.8, text opacity=1]
                at ([xshift=-8pt,yshift=3pt]img.south west) {\footnotesize (a)};
        \end{tikzpicture}
        \caption{}\label{fig:F1_Pairs} 
    \end{subfigure}

    \vspace{-5.5ex}

    \begin{subfigure}{\columnwidth}
        \begin{tikzpicture}
            \node[inner sep=0] (img){\includegraphics[width=\columnwidth]{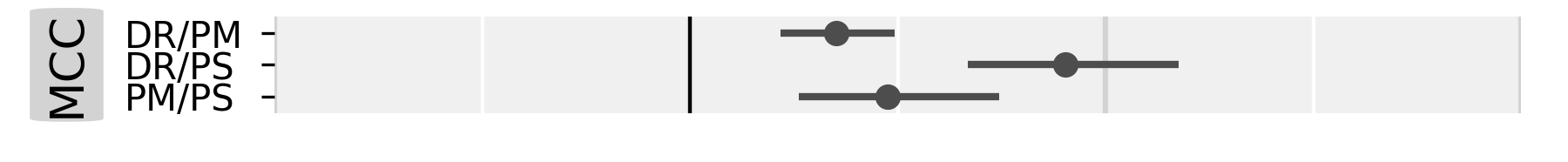}};
            \node[anchor=south west, inner sep=1pt, fill=white, fill opacity=.8, text opacity=1]
                at ([xshift=-8pt,yshift=3pt]img.south west) {\footnotesize (b)};
        \end{tikzpicture}
        \caption{}\label{fig:MCC_Pairs}
    \end{subfigure}

    \vspace{-5.5ex}

    \begin{subfigure}{\columnwidth}
        \begin{tikzpicture}
            \node[inner sep=0] (img){\includegraphics[width=\columnwidth]{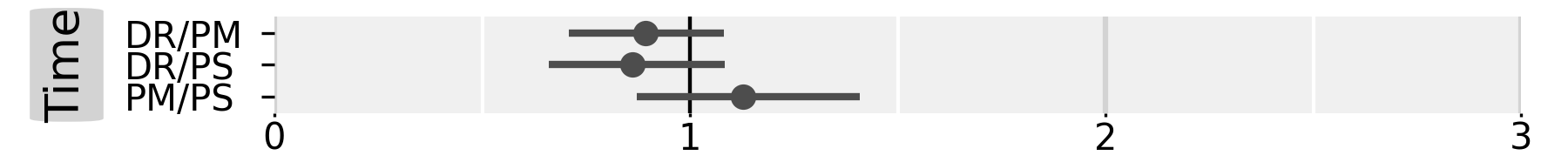}};
            \node[anchor=south west, inner sep=1pt, fill=white, fill opacity=.8, text opacity=1]
                at ([xshift=-8pt,yshift=3pt]img.south west) {\footnotesize (c)};
        \end{tikzpicture}
        \caption{}\label{fig:Time_Pairs}
    \end{subfigure}
    \vspace{-7ex}
    \caption{Pairwise ratio of (a) F1, (b) MCC, and (c) time, 95\% CIs.}
    \label{fig:Pairs}
\end{figure}

In \autoref{fig:Mean} we show the mean completion times and two accuracy metrics \lixiang{(F1 and MCC)} with 95\% confidence intervals across density field conditions \lixiang{(Precomputed Single-resolution (PS), Precomputed Multi-resolution (PM) and Dynamic Resolution (DR))}. We further provide the pairwise ratio with 95\% confidence intervals in \autoref{fig:Pairs}. 
The numerical values of average task completion times, two accuracy scores for each dataset and technique, and the numerical values of pairwise ratio are provided in \autoref{appendix-result_userstudy}.
We now report statistical results in relation to our hypotheses.

\noindent\lixiang{\noindent\textbf{Accuracy.}} For F1, a score of 1 indicates perfect performance and 0 the worst.
For MCC, a score of 1 indicates perfect performance, while $-1$ represents the worst performance. As shown by the mean F1 scores (\FigSubref{fig:Mean}{fig:F1_Mean}) and MCC scores (\FigSubref{fig:Mean}{fig:MCC_Mean}), DR outperformed both PM and PS in accuracy. The pairwise comparisons reveals that DR achieved 1.21--1.44\texttimes{} higher F1 score and 1.22--1.49\texttimes{} higher MCC than PM, and 1.58--1.92\texttimes{} higher F1 and 1.67--2.18\texttimes{} higher MCC than PS. \textbf{H1 is supported}. In addition, PM outperformed PS in both mean F1 and MCC. The pairwise comparisons show that PM achieved 1.21--1.53\texttimes{} higher F1 and 1.26--1.75\texttimes{} higher MCC than PS. Thus, \textbf{H2 is also supported}.
  
\noindent\lixiang{\textbf{Completion Time.}} \FigSubref{fig:Mean}{fig:Time_Mean} shows that selections with DR required less completion time than PM and PS. The pairwise comparisons indicate that DR took 0.70--1.08\texttimes{} as long as PM and 0.66--1.08\texttimes{} as long as PS, suggesting that most of participants completed the tasks more quickly with DR. In a few cases, ratios were only marginally above 1.0, reflecting participants who finished marginally faster with PM or PS. We made related observations during the study: 4 participants ended the tasks early once they realized that they could not achieve satisfactory outcomes with PS or PM. These behaviors, however, do not affect the overall result. We confirm that \textbf{H3 is supported}.
Our results also showed that PM had a slightly shorter mean completion time than PS. The pairwise comparisons indicate that PS took 0.87--1.40\texttimes{} longer than PM, suggesting that the advantage of PM over PS was modest rather than substantial. Thus, \textbf{H4 is not fully supported}.   
  
\begin{figure}[t]
    \centering
    \includegraphics[width=\columnwidth]{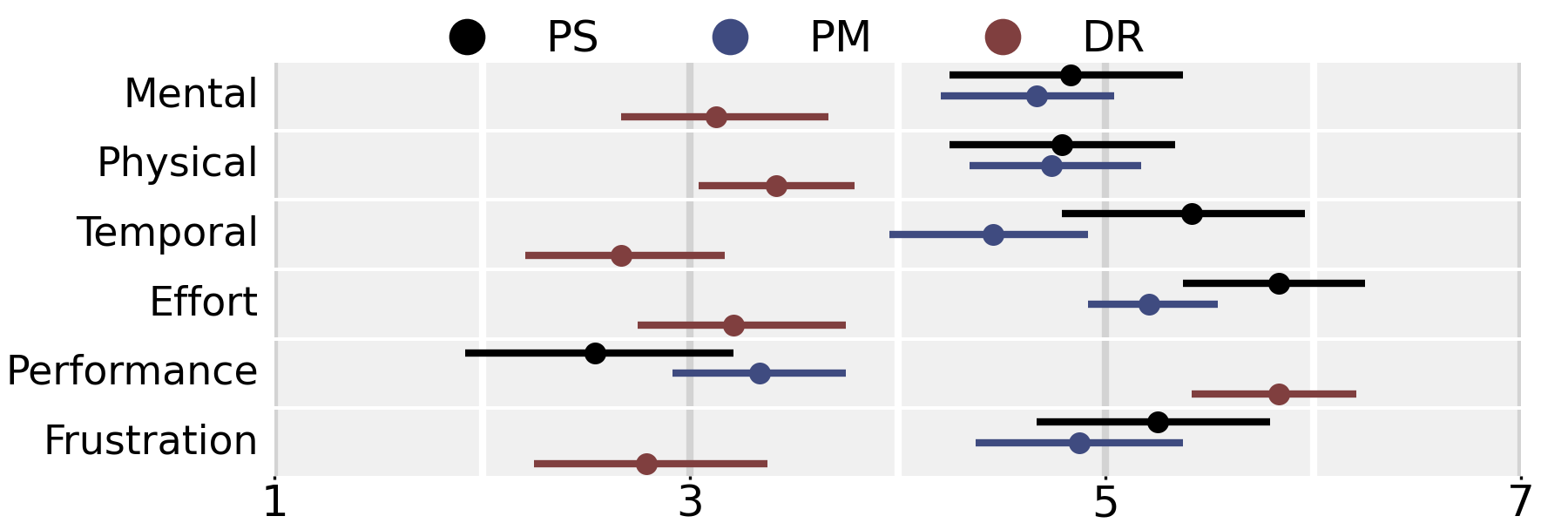}
    \vspace{-1.5em}
    \caption{User workload and performance. Error bars: 95\% CIs.}
    \label{fig:tlx}
\end{figure}

\begin{figure}[t]
    \centering
    \includegraphics[width=\columnwidth]{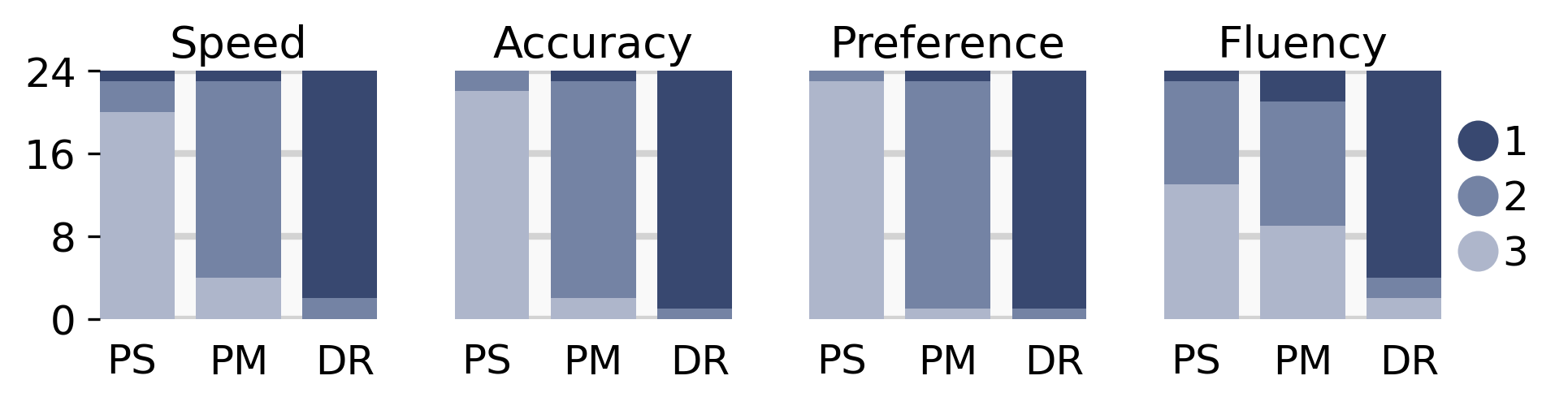}
     \vspace{-1.5em}       
    \caption{Participant rankings on task performance and preference.}
    \label{fig:rank}
\end{figure}

\noindent\lixiang{\textbf{Workload and Preference.}} As shown in \autoref{fig:tlx}, participants reported greater efficiency, lower frustration and time pressure, and reduced mental and physical effort with DR compared to PM and PS. 
Moreover, they ranked the conditions consistently ($DR>PM>PS$) across speed, accuracy, overall preference, and fluency (\autoref{fig:rank}). Thus, \textbf{H5 and H6 are supported}.

\noindent\lixiang{\textbf{Interview.}} In follow-up interviews, we asked participants whether they perceived delays during scale changes (where recomputation happens at the end of scale change) and selections across the three techniques. 
Most participants ($n=20$) reported no noticeable recomputation delay with DR, while four noted slight delays only when the number of points in view was very large. These delays, however, did not affect their navigation experience. 
In contrast, a majority of participants ($n=18$) identified PM as producing the most prominent lag: when using MeTAPoint to select large regions and generate extensive selection volumes, they observed noticeable drops frame rate.
This performance degradation stems from the underlying characteristics of PM: higher-resolution grids increase the number of triangles needed to construct the selection volume and impose heavier computational demands when determining selection thresholds.
These findings highlight scenarios where dynamic density fields offer substantial advantages over precomputed higher-resolution fields, particularly for sparse or large-scale datasets. In such cases, precomputed fields must rely on densely sampled grids to capture comparable detail, whereas dynamic estimation adapts density computation on demand, enabling the capture of fine-grained structures while maintaining smooth frame rates.

\noindent\lingyun{\textbf{Summary.} Improvements in F1 and MCC (\textbf{H1}) and reduced completion time (\textbf{H3}) show that our dynamic KDE improves both selection accuracy and efficiency over precomputed methods. By maintaining scale-accurate density fields, it enables high-fidelity selection with less effort and lower mental demand (\textbf{H5}), and was therefore preferred by participants (\textbf{H6}). These results highlight that on-demand recomputation is essential for precise immersive interaction, as precomputed methods often miss fine-scale details and require additional adjustments.}

\section{Discussion}
\label{sec:discussion}
In this section, we discuss how \method extends to support broader tasks, analyze the trade-offs between dynamic and precomputed density fields, and limitations and future directions.

\subsection{Extending \method}
Point cloud data often lacks a natural scale. Such datasets consist of unstructured points scattered unevenly across scales, sometimes with vast empty ranges, as in astronomical simulations. As a result, interaction and visualization approaches must help users orient themselves within scales and navigate across them. Such tasks require effective methods to compute key features, such as density, as the data changes---forming the starting point of our work.
Multiscale data often contains fine details embedded within broader structures, creating large gaps between scales. A key feature of our dynamic KDE is that it allows users to explore subtle or complex structures on demand, with control remaining in the users' hands. For example, by dynamically adjusting resolution based on the interaction history and on the local data distribution, the navigation approach can guide users toward scale levels where meaningful features emerge. A future direction of research, however, is to determine how our KDE method can be adapted to better handle challenging cases such as if the data distribution dynamically changes across scales. 

\subsection{Trade-offs}

We conducted performance experiments to evaluate \method, comparing its execution time against density estimation on single-core and multi-core CPUs. We also compared selection techniques using \method with those based on precomputed density fields in terms of accuracy and efficiency. These experiments highlight the trade-offs between dynamic and precomputed approaches, offering guidance for future researchers in selecting suitable solutions.

When designing an appropriate strategy, factors such as data scale, hardware capacity, and algorithmic complexity must be considered. Our results show that dynamic KDE consistently delivered higher accuracy and faster computation than precomputed density fields. We also found that multi-resolution density fields achieved better accuracy than single-resolution ones, though with only marginal gains in speed. This suggests that for datasets of moderate scale, precomputing a few density fields may be sufficient---especially when there is only limited memory available---since users can quickly access and switch values without recomputation. For datasets spanning multiple scales, however, dynamic KDE offers clear advantages by computing density fields on demand and supporting seamless multiscale exploration---yet these advantages come with a higher demand for GPU support and GPU\discretionary{/}{}{/}CPU memory and a more complex implementation.
Still, an effective strategy is needed to determine when recomputation should occur, as users may notice delays during density field updates. To address this point, we trigger recomputation after users transition to a new scale. While users take time to observe the data at this scale, the system prepares the density field on the GPU, ensuring readiness for subsequent operations.

\subsection{Limitations and Future Work}

Several limitations remain that open avenues for future work. 

First, although DR dynamically recomputes the density field at a fixed resolution of $64^3$, fine details may still be missed. Although progressive navigation can reveal these features step by step at finer levels, it remains ineffective. Future work could adopt feature-aware resolution schemes that adapt to the underlying distribution, ensuring salient structures in the region of interest are faithfully represented.

Second, our dynamic KDE can be further optimized. Currently, recomputation is triggered even for minor scale changes, leading to unnecessary overhead during frequent zooming. To address this issue, we can introduce a threshold for scale changes, ensuring that recomputation occurs only when the change exceeds a meaningful level. This improvement would reduce redundant updates while maintaining responsiveness. Similarly, we could only compute the KDE for subsets of the dataset at any given zoom level (as opposed to the whole data space), such as the size of the viewport plus some margin around it. Then the recomputation for other regions of the data would be triggered when the user in their interactive navigation gets close to the boundary of the currently computed density field.

\lixiang{Third}, the main time cost of the current recomputation delay in our selection technique lies in transferring the computed density field from the GPU back to the CPU. Future implementations could eliminate this bottleneck by migrating the selection algorithm, or any other interaction entirely onto the GPU, thereby avoiding density transfer from GPU to CPU.

\lixiang{Finally, this work is driven by the demands of immersive multiscale point cloud exploration and therefore focuses on GPU-parallel implementation and system-level optimization of adaptive KDE, rather than proposing new density estimation formulations. While recent KDE methods~\cite{chan:2025:LSS,olsen:2024:TGA} offer improved theoretical efficiency and memory usage, their integration into a real-time interactive VR pipeline remains an important direction for future work.}

\section{Conclusion}
\label{sec:conclusion}

In this work, we present a GPU-accelerated adaptive KDE technique that dynamically recomputes density fields for interactive multiscale exploration. Through two use cases---navigation and selection---we demonstrate that, as an independent module, our \method can be directly integrated into diverse pipelines requiring high-performance density estimation in large-scale scientific simulations.  

Beyond its technical optimization, our \method allows us to explore and pursue a novel interaction paradigm for immersive data exploration. Rather than requiring users to learn dedicated 3D navigation techniques---a long-standing challenge due to the complexity and inherent imprecision of 3D interaction---users essentially simply issue commands akin to ``\emph{show me that in more detail}.'' The structure- and context-aware selection paradigm that we paired with the rapid KDE enabled through our \method approach allows users to simply indicate desired regions of interest to steer their exploration. They can then zoom in to reveal fine-scale detail or adjust their target to focus on another part of the data---ultimately fluidly crossing the scales to find the most informative perspective.

\acknowledgments{
Lingyun Yu was partially supported by the National Natural Science Foundation of China, grant \textnumero\ 62272396.}

\section*{Additional Material Pointers}

We share our additional materials (study results/data and analysis scripts, videos) at \href{https://osf.io/hfu6e}{\texttt{osf.io/hfu6e}}. We also make our prototypical implementation of the \method available at the github repository \href{https://github.com/LixiangZhao98/ScaleFree}{\texttt{github.com\discretionary{/}{}{/}LixiangZhao98\discretionary{/}{}{/}ScaleFree}}.

\section*{Images/graphs/plots/tables/data license/copyright}
We as authors state that all of our figures, graphs, plots, and data tables in this article are and remain under our own personal copyright, with the permission to be used here. We also make them available under the \href{https://creativecommons.org/licenses/by/4.0/}{Creative Commons At\-tri\-bu\-tion 4.0 International (\ccLogo\,\ccAttribution\ \mbox{CC BY 4.0})} license and share them at \href{https://osf.io/hfu6e}{\texttt{osf.io/hfu6e}}.

\bibliographystyle{abbrv-doi-hyperref}

\bibliography{abbreviations,template}

@string { SiggraphCG = "ACM SIGGRAPH Computer Graphics" }

@string { SiggraphCG = "ACM SIGGRAPH Comput Graph" }

@string { TVCG = "IEEE Transactions on Visualization and Computer Graphics" }

@string { TVCG = "IEEE Trans Vis Comput Graph" }

@string { JV = "Journal of Visualization" }

@string { JV = "J Vis" }

@string { VC = "The Visual Computer" }

@string { VC = "Vis Comput" }

@string { Inf = "Information" }

@string { Inf = "Inf" }

@string { VR = "Virtual Reality" }

@string { VR = "Virtual Real" }

@string { BJP = "British Journal of Psychology" }

@string { BJP = "Br J Psychol" }

@string { IEEECompSoc = "IEEE Computer Society" }

@string { IEEECompSoc = "IEEE CS" }

@string { IEEECompSoc = "IEEE Comp.\ Soc." }

@article{Yang:2021:ENI,
  author = {Yang, Yalong and Cordeil, Maxime and Beyer, Johanna and Dwyer, Tim and Marriott, Kim and Pfister, Hanspeter},
  title = {Embodied Navigation in Immersive Abstract Data Visualization: Is Overview+Detail or Zooming Better for {3D} Scatterplots?},
  journal = tvcg,
  year = {2021},
  volume = {27},
  number = {2},
  pages = {1214--1224},
  doi = {10/ghgt52},
  longdoi = {10.1109/TVCG.2020.3030427},
}

@article{Hsu:2013:AMC,
  author = {Hsu, Wei-Hsien and Zhang, Yubo and Ma, Kwan-Liu},
  title = {A Multi-Criteria Approach to Camera Motion Design for Volume Data Animation},
  journal = tvcg,
  year = {2013},
  volume = {19},
  number = {12},
  pages = {2792--2801},
  doi = {10/f5h3m3},
  longdoi = {10.1109/TVCG.2013.123},
}

@inproceedings{garcia:2016:PFU,
  author = {Garc{\'\i}a-Hern{\'a}ndez, Rub{\'e}n Jes{\'u}s and Anthes, Christoph and Wiedemann, Markus and Kranzlm{\"u}ller, Dieter},
  title = {Perspectives for using virtual reality to extend visual data mining in information visualization},
  booktitle = {Proc.\ Aerospace Conf.},
  longbooktitle = {Proc.\ Aerospace Conf.},
  year = {2016},
  publisher = ieeecompsoc,
  address = {Los Alamitos},
  pages = {1151--1161},
  doi = {10/gtmkhw},
  longdoi = {},
}

@inproceedings{whitlock:2020:GPF,
  author = {Whitlock, Matt and Smart, Stephen and Szafir, Danielle Albers},
  title = {Graphical perception for immersive analytics},
  booktitle = {Proc.\ VR},
  longbooktitle = {Proc.\ VR},
  year = {2020},
  publisher = ieeecompsoc,
  address = {Los Alamitos},
  pages = {616--625},
  doi = {10/gktm78},
  longdoi = {10.1109/VR46266.2020.00084},
}

@article{sousa:2009:HMD,
  author = {Sousa Santos, Beatriz and Dias, Paulo and Pimentel, Angela and Baggerman, Jan-Willem and Ferreira, Carlos and Silva, Samuel and Madeira, Joaquim},
  title = {Head-mounted display versus desktop for {3D} navigation in virtual reality: A user study},
  journal = {Multimedia Tools Appl},
  longjournal = {Multimedia tools and applications},
  year = {2009},
  volume = {41},
  number = {1},
  pages = {161--181},
  doi = {10/ccq4dq},
  longdoi = {10.1007/s11042-008-0223-2},
  optpublisher = {Springer},
}

@article{bueckle:2021:3vr,
  author = {Bueckle, Andreas and Buehling, Kilian and Shih, Patrick C. and B{\"o}rner, Katy},
  title = {{3D} virtual reality vs. {2D} desktop registration user interface comparison},
  journal = {PloS one},
  longjournal = {PloS one},
  year = {2021},
  volume = {16},
  number = {10},
  pages = {e0258103},
  doi = {10/qmkr},
  longdoi = {},
  optpublisher = {Public Library of Science San Francisco, CA USA},
}

@article{backurs:2019:SAT,
  author = {Backurs, Arturs and Indyk, Piotr and Wagner, Tal},
  title = {Space and time efficient kernel density estimation in high dimensions},
  journal = {Adv Neural Inf Process Syst},
  longjournal = {Advances in Neural Information Processing Systems},
  year = {2019},
  volume = {32},
  number = {},
  OPTpages = {\textcolor{red}{CHECK/ADD}},
	articleno = {9251},
	numpages = {10},
  doi = {},
  longdoi = {},
  url = {https://papers.nips.cc/paper/2019/file/a2ce8f1706e52936dfad516c23904e3e-Paper.pdf},
  note = {url: \href{https://papers.nips.cc/paper_files/paper/2019/hash/a2ce8f1706e52936dfad516c23904e3e-Abstract.html}{\texttt{papers\discretionary{}{.}{.}nips\discretionary{}{.}{.}cc\discretionary{/}{}{/}paper\_files\discretionary{/}{}{/}paper\discretionary{/}{}{/}2019\discretionary{/}{}{/}hash\discretionary{/}{}{/}a2ce8f1706e52936dfad516c23904e3e\discretionary{}{-}{-}Abstract\discretionary{}{.}{.}html}}},
}

@inproceedings{lampe:2011:IVO,
  author = {Lampe, Ove Daae and Hauser, Helwig},
  title = {Interactive visualization of streaming data with kernel density estimation},
  booktitle = {Proc.\ Pacific Vis},
  longbooktitle = {Proc.\ Pacific Vis},
  year = {2011},
  publisher = ieeecompsoc,
  address = {Los Alamitos},
  pages = {171--178},
  doi = {10/c866v9},
  longdoi = {10.1109/PACIFICVIS.2011.5742387},
}

@inproceedings{chan:2020:quad,
  author = {Chan, Tsz Nam and Cheng, Reynold and Yiu, Man Lung},
  title = {{QUAD}: Quadratic-bound-based kernel density visualization},
  booktitle = {Proc.\ SIGMOD},
  longbooktitle = {Proc.\ SIGMOD},
  year = {2020},
  publisher = {ACM},
  address = {New York},
  pages = {35--50},
  doi = {10/gkpt7c},
  longdoi = {10.1145/3318464.3380561},
}

@article{laha:2012:EOI,
  author = {Laha, Bireswar and Sensharma, Kriti and Schiffbauer, James D. and Bowman, Doug A.},
  title = {Effects of immersion on visual analysis of volume data},
  optjournal = {IEEE transactions on visualization and computer graphics},
  journal = tvcg,
  year = {2012},
  volume = {18},
  number = {4},
  pages = {597--606},
  doi = {10/fx9b5g},
  longdoi = {10.1109/TVCG.2012.42},
  optpublisher = {IEEE},
}

@article{Gansner:2005:TFV,
  author = {Gansner, E.R. and Koren, Y. and North, S.C.},
  title = {Topological fisheye views for visualizing large graphs},
  journal = tvcg,
  longjournal = {IEEE Transactions on Visualization and Computer Graphics},
  year = {2005},
  volume = {11},
  number = {4},
  pages = {457--468},
  doi = {10/cgmnfg},
  longdoi = {10.1109/TVCG.2005.66},
}

@inproceedings{Chheang:2024:AVE,
  author = {Chheang, Vuthea and Weston, Brian Thomas and Cerda, Robert William and Au, Brian and Giera, Brian and Bremer, Peer-Timo and Miao, Haichao},
  title = {A Virtual Environment for Collaborative Inspection in Additive Manufacturing},
  booktitle = {Proc.\ CHI},
  longbooktitle = {Extended Abstracts of the CHI Conference on Human Factors in Computing Systems},
  year = {2024},
  publisher = {ACM},
  address = {New York},
  articleno = {26},
  numpages = {7},
  pages = {26:1--26:7},
  doi = {10/g92v9w},
  longdoi = {10.1145/3613905.3650730},
  url = {https://doi.org/10.1145/3613905.3650730},
  optisbn = {9798400703317},
  optlocation = {Honolulu, HI, USA},
  optseries = {CHI EA '24},
}

@article{Lorensen:1987:MCH,
  author = {Lorensen, William E. and Cline, Harvey E.},
  title = {Marching Cubes: A High Resolution {3D} Surface Construction Algorithm},
  journal = siggraphcg,
  year = {1987},
  optmonth = aug,
  volume = {21},
  number = {4},
  pages = {163--169},
  doi = {10/ft9gsh},
  longdoi = {10.1145/37402.37422},
}

@misc{htcvivepro2,
  author = {Rory Brown},
  title = {{HTC} Vive Pro2},
  howpublished = {Web site: \href{https://vr-compare.com/headset/htcvivepro2}{\texttt{{vr\discretionary{}{-}{-}compare\discretionary{.}{}{.}com\discretionary{/}{}{/}headset\discretionary{/}{}{/}htcvivepro2}}}},
  note = {Last accessed: Sep.\ 2025},
}

@article{Wyvill:1986:DSS,
  author = {Wyvill, G. and McPheeters, C. and Wyvill, B.},
  title = {Data structure for soft objects},
  journal = vc,
  year = {1986},
  volume = {2},
  number = {4},
  pages = {227--234},
  doi = {10/dndmwc},
  longdoi = {10.1007/BF01900346},
}

@inproceedings{Klacansky:2022:VIA,
  author = {Klacansky, Pavol and Miao, Haichao and Gyulassy, Attila and Townsend, Andrew and Champley, Kyle and Tringe, Joseph and Pascucci, Valerio and Bremer, Peer-Timo},
  title = {Virtual Inspection of Additively Manufactured Parts},
  booktitle = {Proc.\ PacificVis},
  longbooktitle = {2022 IEEE 15th Pacific Visualization Symposium (PacificVis)},
  year = {2022},
  publisher = ieeecompsoc,
  address = {Los Alamitos},
  volume = {},
  number = {},
  pages = {81--90},
  doi = {10/g92v9x},
  longdoi = {10.1109/PacificVis53943.2022.00017},
}

@article{Halladjian:2022:MUI,
  author = {Halladjian, Sarkis and Kouřil, David and Miao, Haichao and Gröller, M. Eduard and Viola, Ivan and Isenberg, Tobias},
  title = {Multiscale Unfolding: Illustratively Visualizing the Whole Genome at a Glance},
  journal = tvcg,
  longjournal = {IEEE Transactions on Visualization and Computer Graphics},
  year = {2022},
  volume = {28},
  number = {10},
  pages = {3456--3470},
  doi = {10/kt3m},
  longdoi = {10.1109/TVCG.2021.3065443},
}

@article{Cho:2018:MS7,
  author = {Cho, Isaac and Li, Jialei and Wartell, Zachary},
  title = {Multi-Scale {7DOF} View Adjustment},
  journal = tvcg,
  longjournal = {IEEE Transactions on Visualization and Computer Graphics},
  year = {2018},
  volume = {24},
  number = {3},
  pages = {1331--1344},
  doi = {10/gcx789},
  longdoi = {10.1109/TVCG.2017.2668405},
}

@inproceedings{LaViola:2001:HFM,
  author = {LaViola, Joseph J. and Feliz, Daniel Acevedo and Keefe, Daniel F. and Zeleznik, Robert C.},
  title = {Hands-free multi-scale navigation in virtual environments},
  booktitle = {Proc.\ I3D},
  longbooktitle = {Proceedings of the 2001 Symposium on Interactive 3D Graphics},
  year = {2001},
  publisher = {ACM},
  address = {New York},
  optnumpages = {7},
  pages = {9--15},
  doi = {10/bdmkfc},
  longdoi = {10.1145/364338.364339},
  url = {https://doi.org/10.1145/364338.364339},
  optisbn = {1581132921},
  optseries = {I3D '01},
}

@inproceedings{Colin:1997:CSF,
  author = {Ware, Colin and Fleet, Daniel},
  title = {Context sensitive flying interface},
  booktitle = {Proc.\ I3D},
  longbooktitle = {Proceedings of the 1997 Symposium on Interactive 3D Graphics},
  year = {1997},
  publisher = {ACM},
  address = {New York},
  pages = {127--130},
  doi = {10/drbfvq},
  longdoi = {10.1145/253284.253319},
  url = {https://doi.org/10.1145/253284.253319},
  optisbn = {0897918843},
  optlocation = {Providence, Rhode Island, USA},
  optseries = {I3D '97},
}

@article{Kouvril:2021:HyperLabels,
  author = {Kou\v{r}il, David and Isenberg, Tobias and Kozl\'{\i}kov\'{a}, Barbora and Meyer, Miriah and Gr\"{o}ller, M. Eduard and Viola, Ivan},
  title = {Hyper{L}abels: Browsing of Dense and Hierarchical Molecular {3D} Models},
  journal = tvcg,
  longjournal = {IEEE Transactions on Visualization and Computer Graphics},
  year = {2021},
  optmonth = aug,
  optaddress = {USA},
  volume = {27},
  number = {8},
  optnumpages = {12},
  pages = {3493--3504},
  doi = {10/kt4m},
  longdoi = {10.1109/TVCG.2020.2975583},
  url = {https://doi.org/10.1109/TVCG.2020.2975583},
  optissn = {1077-2626},
  optissue_date = {Aug. 2021},
  optpublisher = {IEEE Educational Activities Department},
}

@article{Song:1994:LIL,
  author = {Deyang Song and Norman, M.I.},
  title = {Looking in, looking out: Exploring multiscale data with virtual reality},
  journal = {IEEE Comput Sci Eng},
  longjournal = {IEEE Computational Science and Engineering},
  year = {1994},
  volume = {1},
  number = {3},
  pages = {53--64},
  doi = {10/csw826},
  longdoi = {10.1109/MCSE.1994.313168},
}

@inproceedings{Fu:2010:MTT,
  author = {Fu, Chi-Wing and Goh, Wooi-Boon and Ng, Junxiang Allen},
  title = {Multi-touch techniques for exploring large-scale {3D} astrophysical simulations},
  booktitle = {Proc.\ CHI},
  longbooktitle = {Proceedings of the SIGCHI Conference on Human Factors in Computing Systems},
  year = {2010},
  publisher = {ACM},
  address = {New York},
  optnumpages = {10},
  pages = {2213--2222},
  doi = {10/drxzf7},
  longdoi = {10.1145/1753326.1753661},
  url = {https://doi.org/10.1145/1753326.1753661},
  optisbn = {9781605589299},
  optlocation = {Atlanta, Georgia, USA},
  optseries = {CHI '10},
}

@article{Springel:2005:SimulationsOT,
  author = {Volker Springel and Simon D. M. White and Adrian Jenkins and Carlos S. Frenk and Naoki Yoshida and Liang Gao and Julio F. Navarro and Robert J. Thacker and Darren J. Croton and John C. Helly and John A. Peacock and Shaun Cole and Peter A. Thomas and Hugh M. P. Couchman and August E. Evrard and J. M. Colberg and Frazer R. Pearce},
  title = {Simulations of the formation, evolution and clustering of galaxies and quasars},
  journal = {Nature},
  longjournal = {Nature},
  year = {2005},
  volume = {435},
  number = {},
  pages = {629--636},
  doi = {10/c3cmxr},
  longdoi = {10.1038/nature03597},
}

@article{hahn:2007:PDM,
  author = {Hahn, Oliver and Porciani, Cristiano and Carollo, C. Marcella and Dekel, Avishai},
  title = {Properties of dark matter haloes in clusters, filaments, sheets and voids},
  journal = {Mon Not R Astron Soc},
  longjournal = {Monthly Notices of the Royal Astronomical Society},
  year = {2007},
  volume = {375},
  number = {2},
  pages = {489--499},
  doi = {10/dxbdjj},
  longdoi = {10.1111/j.1365-2966.2006.11318.x},
  optpublisher = {Blackwell Publishing Ltd Oxford, UK},
}

@article{olsen:2024:TGA,
  author = {Olsen, Kenny and Lindrup, Rasmus M. Hoeegh and M{\o}rup, Morten},
  title = {Think Global, Adapt Local: Learning Locally Adaptive K-Nearest Neighbor Kernel Density Estimators},
  journal = {PMLR},
  longjournal = {PMLR},
  year = {2024},
  volume = {238},
  number = {},
  pages = {4114--4122},
  doi = {},
  longdoi = {},
  url = {https://proceedings.mlr.press/v238/olsen24a.html},
  note = {url: \href{https://proceedings.mlr.press/v238/olsen24a.html}{\texttt{proceedings\discretionary{}{.}{.}mlr\discretionary{}{.}{.}press\discretionary{/}{}{/}v238\discretionary{/}{}{/}olsen24a\discretionary{}{.}{.}html}}},
}

@article{springel:2008:aquarius,
  author = {Springel, Volker and Wang, Jie and Vogelsberger, Mark and Ludlow, Aaron and Jenkins, Adrian and Helmi, Amina and Navarro, Julio F. and Frenk, Carlos S. and White, Simon DM},
  title = {The Aquarius Project: The subhaloes of galactic haloes},
  journal = {Mon Not R Astron Soc},
  longjournal = {Monthly Notices of the Royal Astronomical Society},
  year = {2008},
  volume = {391},
  number = {4},
  pages = {1685--1711},
  doi = {10/fsjgzw},
  longdoi = {10.1111/j.1365-2966.2008.14066.x},
  optpublisher = {Blackwell Publishing Ltd Oxford, UK},
}

@article{Yu:2016:CEE,
  author = {Lingyun Yu and Konstantinos Efstathiou and Petra Isenberg and Tobias Isenberg},
  title = {{CAST}: Effective and Efficient User Interaction for Context-Aware Selection in {3D} Particle Clouds},
  journal = tvcg,
  year = {2016},
  optmonth = jan,
  volume = {22},
  number = {1},
  pages = {886--895},
  doi = {10/kt5n},
  longdoi = {10.1109/TVCG.2015.2467202},
  optnote = {\href{https://hal.archives-ouvertes.fr/hal-01178051}{\oatext}},
}

@inproceedings{prouzeau:2019:scaptics,
  author = {Prouzeau, Arnaud and Cordeil, Maxime and Robin, Clement and Ens, Barrett and Thomas, Bruce H. and Dwyer, Tim},
  title = {Scaptics and highlight-planes: Immersive interaction techniques for finding occluded features in {3D} scatterplots},
  booktitle = {Proc.\ CHI},
  longbooktitle = {Proceedings of the 2019 CHI Conference on Human Factors in Computing Systems},
  year = {2019},
  publisher = {ACM},
  address = {New York},
  OPTpages = {1--12},
	articleno = {325},
	numpages = {12},
  doi = {10/gnkk76},
  longdoi = {10.1145/3290605.3300555},
}

@article{Yu:2012:ESA,
  author = {Yu, Lingyun and Efstathiou, Konstantinos and Isenberg, Petra and Isenberg, Tobias},
  title = {Efficient Structure-Aware Selection Techniques for {3D} Point Cloud Visualizations with {2DOF} Input},
  journal = tvcg,
  year = {2012},
  volume = {18},
  number = {12},
  pages = {2245--2254},
  doi = {10/f4fv9z},
  longdoi = {10.1109/TVCG.2012.217},
}

@article{zhao:2023:metacast,
  author = {Zhao, Lixiang and Isenberg, Tobias and Xie, Fuqi and Liang, Hai-Ning and Yu, Lingyun},
  title = {Me{TACAST}: Target-and Context-aware Spatial Selection in {VR}},
  journal = tvcg,
  year = {2024},
  volume = {30},
  number = {1},
  pages = {480--494},
  doi = {10/gtnn25},
  longdoi = {10.1109/TVCG.2023.3326517},
}

@article{diggle:1985:KDEDiggle,
  author = {Diggle, Peter},
  title = {A kernel method for smoothing point process data},
  journal = {J R Stat Soc C (Appl Stat)},
  longjournal = {Journal of the Royal Statistical Society: Series C (Applied Statistics)},
  year = {1985},
  volume = {34},
  number = {2},
  pages = {138--147},
  doi = {10/bsh8ht},
  longdoi = {10.2307/2347366},
  optpublisher = {Wiley Online Library},
}

@article{ferdosi:2011:CDE,
  author = {Ferdosi, BJ and Buddelmeijer, Hugo and Trager, SC and Wilkinson, MHF and Roerdink, JBTM},
  title = {Comparison of density estimation methods for astronomical datasets},
  journal = {Astron Astrophys},
  longjournal = {Astronomy \& Astrophysics},
  year = {2011},
  volume = {531},
  number = {},
  articleno = {A114},
  numpages = {16},
  pages = {A114:1--A114:16},
  doi = {10/cj3s5m},
  longdoi = {10.1051/0004-6361/201116878},
  optpublisher = {EDP Sciences},
}

@article{wilkinson:1995:dataplot,
  author = {Wilkinson, M. H. F. and Meijer, B. C.},
  title = {{DATAPLOT}: A graphical display package for bacterial morphometry and fluorimetry data},
  journal = {Comput Methods Programs Biomed},
  longjournal = {Computer Methods and Programs in Biomedicine},
  year = {1995},
  volume = {47},
  number = {1},
  pages = {35--49},
  doi = {10/dnr53p},
  longdoi = {10.1016/0169-2607(95)01628-7},
  optpublisher = {Elsevier},
}

@techreport{kakde:2005:kdtree,
  author = {H. M. Kakde},
  title = {Range Searching using {KD} Tree},
  institution = {Dept.\ of Computer Science, Florida State Univ.},
  year = {2005},
  doi = {},
  longdoi = {},
  note         = {Available: \url{https://users.cs.utah.edu/\%7Elifeifei/cis5930/kdtree.pdf}}
}

@inproceedings{chan:2025:LSS,
  author = {Chan, Tsz Nam and Ip, Pak Lon and Zhu, Bojian and Wu, Dingming and Xu, Jianliang and Jensen, Christian S. and others},
  title = {Large-Scale Spatiotemporal Kernel Density Visualization},
  booktitle = {Proc.\ ICDE},
  longbooktitle = {2025 IEEE 41st International Conference on Data Engineering (ICDE)},
  year = {2025},
  publisher = ieeecompsoc,
  address = {Los Alamitos},
  pages = {99--113},
  doi = {10/qmw4},
  longdoi = {10.1109/ICDE65448.2025.00015},
}

@inproceedings{stoakley:1995:VRWIM,
  author = {Stoakley, Richard and Conway, Matthew J. and Pausch, Randy},
  title = {Virtual reality on a {WIM}: Interactive worlds in miniature},
  booktitle = {Proc.\ CHI},
  longbooktitle = {Proceedings of the SIGCHI conference on Human factors in computing systems},
  year = {1995},
  publisher = {ACM},
  address = {New York},
  pages = {265--272},
  doi = {10/bx762s},
  longdoi = {10.1145/223904.223938},
}

@inproceedings{wingrave:2006:SSWIM,
  author = {Wingrave, Chadwick A. and Haciahmetoglu, Yonca and Bowman, Doug A.},
  title = {Overcoming world in miniature limitations by a scaled and scrolling {WIM}},
  booktitle = {Proc.\ 3DUI},
  longbooktitle = {3D User Interfaces (3DUI'06)},
  year = {2006},
  publisher = ieeecompsoc,
  address = {Los Alamitos},
  pages = {11--16},
  doi = {10/bbp9k4},
  longdoi = {10.1109/VR.2006.106},
}

@inproceedings{zhao:2022:Lwim,
  author = {Zhao, Lixiang and Cao, Nieyu and He, Shuqi and Liang, Hai-Ning and Yu, Lingyun},
  title = {L-WiM: Collaborative exploration in immersive environments},
  booktitle = {ISMAR Adjunct Proc.},
  longbooktitle = {2022 IEEE International Symposium on Mixed and Augmented Reality Adjunct (ISMAR-Adjunct)},
  year = {2022},
  publisher = ieeecompsoc,
  address = {Los Alamitos},
  pages = {118--123},
  doi = {10/gtmh3d},
  longdoi = {10.1109/ISMAR-Adjunct57072.2022.00031},
}

@article{Baguley:2009:ES,
  author = {Baguley, Thom},
  title = {Standardized or simple effect size: What should be reported?},
  journal = bjp,
  year = {2009},
  optmonth = aug,
  volume = {100},
  number = {3},
  pages = {603--617},
  doi = {10/bnw2nb},
  longdoi = {10.1348/000712608X377117},
  optpublisher = {Wiley Online Library},
}

@article{Cumming:2013:NS,
  author = {Cumming, Geoff},
  title = {The new statistics: Why and how},
  optjournal = {Psychological Science},
  journal = {Psychol Sci},
  longjournal = {Psychol Sci},
  year = {2014},
  optmonth = jan,
  volume = {25},
  number = {1},
  pages = {7--29},
  doi = {10/5k3},
  longdoi = {10.1177/0956797613504966},
  opturl = {http://pss.sagepub.com/content/25/1/7.abstract},
  optissn = {1467-9280},
  pmid = {24220629},
}

@book{VandenBos:2009:PMAPA,
  title = {Publication Manual of the American Psychological Association},
  editor = {VandenBos, Gary R.},
  year = {2009},
  publisher = {APA},
  address = {Washington, DC},
  url = {http://www.apastyle.org/manual/},
  edition = {6\textsuperscript{th}},
  note = {url: \href{http://www.apastyle.org/manual/}{\texttt{apastyle\discretionary{}{.}{.}org\discretionary{/}{}{/}manual}}},
  optpublisher = {American Psychological Association},
}

@incollection{Dragicevic:2016:FSC,
  author = {Dragicevic, Pierre},
  title = {Fair Statistical Communication in {HCI}},
  booktitle = {Modern Statistical Methods for HCI},
  longbooktitle = {Modern Statistical Methods for HCI},
  editor = {Judy Robertson and Maurits Kaptein},
  year = {2016},
  publisher = {Springer},
  address = {Cham},
  chapter = {13},
  pages = {291--330},
  doi = {10/ggd8gc},
  longdoi = {10.1007/978-3-319-26633-6_13},
  opturl = {http://software.mauritskaptein.com/StatisticsForHCI/content-overview/},
  optnote = {In press; pre-print available from the author},
}

@inproceedings{Dragicevic:2014:RAH,
  author = {Dragicevic, Pierre and Chevalier, Fanny and Huot, St\'{e}phane},
  title = {Running an {HCI} Experiment in Multiple Parallel Universes},
  booktitle = {CHI Extended Abstracts},
  longbooktitle = {CHI Extended Abstracts},
  optbooktitle = {Extended Abstracts on Human Factors in Computing Systems},
  year = {2014},
  publisher = {ACM},
  address = {New York},
  pages = {607--618},
  doi = {10/gpcs8n},
  longdoi = {10.1145/2559206.2578881},
}

@inproceedings{pivovar:2022:SWIM,
  author = {Pivovar, Jarod and DeGuzman, Jasmine and Rosenberg, Evan Suma},
  title = {Virtual reality on a {SWIM}: Scalable world in miniature},
  booktitle = {Proc.\ VRW},
  longbooktitle = {2022 IEEE Conference on Virtual Reality and 3D User Interfaces Abstracts and Workshops (VRW)},
  year = {2022},
  publisher = ieeecompsoc,
  address = {Los Alamitos},
  pages = {912--913},
  doi = {10/gt2485},
  longdoi = {10.1109/VRW55335.2022.00307},
}

@inproceedings{kopper:2006:DEN,
  author = {Kopper, Regis and Ni, Tao and Bowman, Doug A. and Pinho, Marcio},
  title = {Design and evaluation of navigation techniques for multiscale virtual environments},
  booktitle = {Proc.\ VR},
  longbooktitle = {Ieee virtual reality conference (vr 2006)},
  year = {2006},
  publisher = ieeecompsoc,
  address = {Los Alamitos},
  pages = {175--182},
  doi = {10/cjs6g6},
  longdoi = {10.1109/VR.2006.47},
}

@article{TLXScale,
  author = {Hart, Sandra},
  title = {Nasa-task load index ({NASA}-{TLX}); {20} years later},
  journal = {Proc Hum Factors Ergon Soc Annu Meet},
  longjournal = {Proc Hum Factors Ergon Soc Annu Meet},
  year = {2006},
  optmonth = oct,
  volume = {50},
  number = {9},
  pages = {904--908},
  doi = {10/fzvtd4},
  longdoi = {10.1177/154193120605000909},
  journalopt = {Proceedings of the Human Factors and Ergonomics Society Annual Meeting},
}

@inproceedings{bacim:2009:WTM,
  author = {Bacim, Felipe and Bowman, Doug and Pinho, Marcio},
  title = {Wayfinding techniques for multiscale virtual environments},
  booktitle = {Proc.\ 3DUI},
  longbooktitle = {2009 IEEE Symposium on 3D User Interfaces},
  year = {2009},
  publisher = ieeecompsoc,
  address = {Los Alamitos},
  pages = {67--74},
  doi = {10/fg3dbh},
  longdoi = {10.1109/3DUI.2009.4811207},
}

@inproceedings{argelaguet:2016:giant,
  author = {Argelaguet, Ferran and Maignant, Morgant},
  title = {Gi{A}nt: Stereoscopic-compliant multi-scale navigation in {VE}s},
  booktitle = {Proc.\ VRST},
  longbooktitle = {Proceedings of the 22nd acm conference on virtual reality software and technology},
  year = {2016},
  publisher = {ACM},
  address = {New York},
  pages = {269--277},
  doi = {10/p46w},
  longdoi = {10.1145/2993369.2993391},
}

@article{weissker:2024:TTF,
  author = {Weissker, Tim and Franzgrote, Matthis and Kuhlen, Torsten},
  title = {Try this for size: Multi-scale teleportation in immersive virtual reality},
  journal = tvcg,
  longjournal = {IEEE Transactions on Visualization and Computer Graphics},
  year = {2024},
  volume = {30},
  number = {5},
  pages = {2298--2308},
  doi = {10/p5ft},
  longdoi = {10.1109/TVCG.2024.3372043},
  optpublisher = {IEEE},
}

@inproceedings{pavanatto:2025:EMN,
  author = {Pavanatto, Leonardo and Giovannelli, Alexander and Giera, Brian and Bremer, Timo and Miao, Haichao and Bowman, Doug A.},
  title = {Exploring Multiscale Navigation of Homogeneous and Dense Objects with Progressive Refinement in Virtual Reality},
  booktitle = {Proc.\ VR},
  longbooktitle = {2025 IEEE Conference Virtual Reality and 3D User Interfaces (VR)},
  year = {2025},
  publisher = ieeecompsoc,
  address = {Los Alamitos},
  pages = {228--237},
  doi = {10/p5fw},
  longdoi = {10.1109/VR59515.2025.00047},
}

@inproceedings{Wills:1996:S5W,
  author = {Wills, Graham J.},
  title = {Selection: {524},{288} ways to say ``this is interesting''},
  booktitle = {Proc.\ InfoVis},
  longbooktitle = {Proc.\ InfoVis},
  year = {1996},
  publisher = ieeecompsoc,
  address = {Los Alamitos},
  pages = {54--60},
  doi = {10/bjq6dh},
  longdoi = {10.1109/INFVIS.1996.559216},
}

@article{besanccon:2021:state-of-art,
  author = {Besan{\c{c}}on, Lonni and Ynnerman, Anders and Keefe, Daniel F. and Yu, Lingyun and Isenberg, Tobias},
  title = {The state of the art of spatial interfaces for {3D} visualization},
  journal = {Comput Graph Forum},
  longjournal = {Computer Graphics Forum},
  year = {2021},
  organization = {Wiley Online Library},
  volume = {40},
  number = {1},
  pages = {293--326},
  doi = {10/gjbpxp},
  longdoi = {10.1111/cgf.14189},
}

@article{chen:2015:CWR,
  author = {Chen, Yen-Chi and Ho, Shirley and Freeman, Peter E. and Genovese, Christopher R. and Wasserman, Larry},
  title = {Cosmic web reconstruction through density ridges: Method and algorithm},
  journal = {Mon Not R Astron Soc},
  longjournal = {Monthly Notices of the Royal Astronomical Society},
  year = {2015},
  volume = {454},
  number = {1},
  pages = {1140--1156},
  doi = {10/f7wrzq},
  longdoi = {10.1093/mnras/stv1996},
  optpublisher = {Oxford University Press},
}

@inproceedings{he:2007:EGA,
  author = {He, Bingsheng and Govindaraju, Naga K. and Luo, Qiong and Smith, Burton},
  title = {Efficient gather and scatter operations on graphics processors},
  booktitle = {Proc.\ SC},
  longbooktitle = {Proceedings of the 2007 ACM/IEEE Conference on Supercomputing},
  year = {2007},
  publisher = {ACM},
  address = {New York},
  articleno = {46},
  numpages = {12},
  pages = {46:1--46:12},
  doi = {10/fgtjzk},
  longdoi = {10.1145/1362622.1362684},
}

@article{cautun:2013:NEXUS,
  author = {Cautun, Marius and van de Weygaert, Rien and Jones, Bernard JT},
  title = {{NEXUS}: Tracing the cosmic web connection},
  journal = {Mon Not R Astron Soc},
  longjournal = {Monthly Notices of the Royal Astronomical Society},
  year = {2013},
  volume = {429},
  number = {2},
  pages = {1286--1308},
  doi = {10/f4wxgz},
  longdoi = {10.1093/mnras/sts416},
  optpublisher = {The Royal Astronomical Society},
}

@article{shivashankar:2015:Felix,
  author = {Shivashankar, Nithin and Pranav, Pratyush and Natarajan, Vijay and van de Weygaert, Rien and Bos, EG Patrick and Rieder, Steven},
  title = {Felix: A topology based framework for visual exploration of cosmic filaments},
  journal = tvcg,
  longjournal = {IEEE Transactions on Visualization and Computer Graphics},
  year = {2016},
  volume = {22},
  number = {6},
  pages = {1745--1759},
  doi = {10/f8q5rz},
  longdoi = {10.1109/TVCG.2015.2452919},
  optpublisher = {IEEE},
}

@article{pfeifer:2022:COWS,
  author = {Pfeifer, Simon and Libeskind, Noam I. and Hoffman, Yehuda and Hellwing, Wojciech A. and Bilicki, Maciej and Naidoo, Krishna},
  title = {{COWS}: A filament finder for Hessian cosmic web identifiers},
  journal = {Mon Not R Astron Soc},
  longjournal = {Monthly Notices of the Royal Astronomical Society},
  year = {2022},
  volume = {514},
  number = {1},
  pages = {470--479},
  doi = {10/p5fx},
  longdoi = {10.1093/mnras/stac1382},
  optpublisher = {Oxford University Press},
}

@book{Silverman:1986:DEF,
  author = {Silverman, B. W.},
  title = {Density Estimation for Statistics and Data Analysis},
  year = {1986},
  publisher = {Chapman and Hall/CRC},
  address = {New York},
  doi = {10/gqbn6x},
  longdoi = {10.1201/9781315140919},
}

@article{brunsdon:1995:EPS,
  author = {Brunsdon, Chris},
  title = {Estimating probability surfaces for geographical point data: An adaptive kernel algorithm},
  journal = {Comput GeoSci},
  longjournal = {Computers \& Geosciences},
  year = {1995},
  volume = {21},
  number = {7},
  pages = {877--894},
  doi = {10/dnt6g8},
  longdoi = {10.1016/0098-3004(95)00020-9},
  optpublisher = {Elsevier},
}

@article{yuan:2020:AFM,
  author = {Yuan, Zunli and Jarvis, Matt J. and Wang, Jiancheng},
  title = {A flexible method for estimating luminosity functions via kernel density estimation},
  journal = {Astrophys J Suppl Ser},
  longjournal = {The Astrophysical Journal Supplement Series},
  year = {2020},
  volume = {248},
  number = {1},
  articleno = {1},
  numpages = {18},
  pages = {1:1--1:18},
  doi = {10/p5qc},
  longdoi = {10.3847/1538-4365/ab855b},
  optpublisher = {IOP Publishing},
}

@article{yuan:2020:AFM:2,
  author = {Zunli Yuan and Xibin Zhang and Jiancheng Wang and Xiangming Cheng and Wenjie Wang},
  title = {A flexible method for estimating luminosity functions via kernel density estimation. {II}. {G}eneralization and Python Implementation},
  journal = {Astrophys J Suppl Ser},
  longjournal = {The Astrophysical Journal Supplement Series},
  year = {2022},
  volume = {260},
  number = {1},
  articleno = {10},
  numpages = {15},
  pages = {10:1--10:15},
  doi = {10/p5f2},
  longdoi = {10.3847/1538-4365/ac596a},
}

@article{pelz:2023:ADB,
  author = {Pelz, Maria-Theresia and Schartau, Markus and Somes, Christopher J. and Lampe, Vanessa and Slawig, Thomas},
  title = {A diffusion-based kernel density estimator (diff{KDE}, version 1) with optimal bandwidth approximation for the analysis of data in geoscience and ecological research},
  journal = {GeoSci Model Dev},
  longjournal = {Geoscientific Model Development},
  year = {2023},
  volume = {16},
  number = {22},
  pages = {6609--6634},
  doi = {10/p5f3},
  longdoi = {10.5194/gmd-16-6609-2023},
  optpublisher = {Copernicus Publications G{\"o}ttingen, Germany},
}

@article{saito:2019:ASA,
  author = {Saito, Kotaro and Yano, Masao and Hino, Hideitsu and Shoji, Tetsuya and Asahara, Akinori and Morita, Hidekazu and Mitsumata, Chiharu and Kohlbrecher, Joachim and Ono, Kanta},
  title = {Accelerating small-angle scattering experiments on anisotropic samples using kernel density estimation},
  journal = {Sci Rep},
  longjournal = {Scientific reports},
  year = {2019},
  volume = {9},
  number = {1},
  articleno = {1526},
  numpages = {10},
  pages = {1526:1--1526:10},
  doi = {10/p5f4},
  longdoi = {10.1038/s41598-018-37345-5},
  optpublisher = {Nature Publishing Group UK London},
}

@article{Zhang:2017:AGA,
  author = {Zhang, Guiming and Zhu, A-Xing and Huang, Qunying},
  title = {A {GPU}-accelerated adaptive kernel density estimation approach for efficient point pattern analysis on spatial big data},
  journal = {Int J Geogr Inf Sci},
  longjournal = {International Journal of Geographical Information Science},
  year = {2017},
  volume = {31},
  number = {10},
  pages = {2068--2097},
  doi = {10/ggfmp5},
  longdoi = {10.1080/13658816.2017.1324975},
  optpublisher = {Taylor \& Francis},
}

@book{burt:2009:ESF,
  author = {Burt, James E. and Barber, Gerald M. and Rigby, David L.},
  title = {Elementary Statistics for Geographers},
  year = {2009},
  publisher = {Guilford Press},
  address = {New York},
  doi = {10/br6ngd},
  longdoi = {10.1111/j.1467-985X.2009.00634_5.x},
  edition = {3\textsuperscript{rd}},
}

@article{fleming:2017:ANK,
  author = {Fleming, Christen H. and Calabrese, Justin M.},
  title = {A new kernel density estimator for accurate home-range and species-range area estimation},
  journal = {Methods Ecol Evol},
  longjournal = {Methods in Ecology and Evolution},
  year = {2017},
  volume = {8},
  number = {5},
  pages = {571--579},
  doi = {10/f98k6z},
  longdoi = {10.1111/2041-210X.12673},
  optpublisher = {Wiley Online Library},
}

@article{xie:2008:KDET,
  author = {Xie, Zhixiao and Yan, Jun},
  title = {Kernel density estimation of traffic accidents in a network space},
  journal = {Comput Env Urban Syst},
  longjournal = {Computers, environment and urban systems},
  year = {2008},
  volume = {32},
  number = {5},
  pages = {396--406},
  doi = {10/b7gxpq},
  longdoi = {10.1016/j.compenvurbsys.2008.05.001},
  optpublisher = {Elsevier},
}

@book{scott:2015:MDE,
  author = {Scott, David W.},
  title = {Multivariate Density Estimation: Theory, Practice, and Visualization},
  year = {2015},
  publisher = {John Wiley \& Sons},
  address = {Hoboken},
  doi = {10/p5f5},
  longdoi = {10.1002/9781118575574},
  edition = {2\textsuperscript{nd}},
}

@article{epanechnikov:1969:Epanechnikov,
  author = {Epanechnikov, Vassiliy A.},
  title = {Non-parametric estimation of a multivariate probability density},
  journal = {Theory Probab Appl},
  longjournal = {Theory of Probability \& Its Applications},
  year = {1969},
  volume = {14},
  number = {1},
  pages = {153--158},
  doi = {10/bjdbsc},
  longdoi = {10.1137/1114019},
  optpublisher = {SIAM},
}

@article{zhang:2016:EPP,
  author = {Zhang, Guiming and Huang, Qunying and Zhu, A-Xing and Keel, John H.},
  title = {Enabling point pattern analysis on spatial big data using cloud computing: Optimizing and accelerating Ripley's K function},
  journal = {Int J Geogr Inf Sci},
  longjournal = {International Journal of Geographical Information Science},
  year = {2016},
  volume = {30},
  number = {11},
  pages = {2230--2252},
  doi = {10/gjkw5j},
  longdoi = {10.1080/13658816.2016.1170836},
  optpublisher = {Taylor \& Francis},
}

@article{Halladjian:2020:ScaleTrotter,
  author = {Halladjian, Sarkis and Miao, Haichao and Kouřil, David and Gröller, M. Eduard and Viola, Ivan and Isenberg, Tobias},
  title = {Scale Trotter: Illustrative Visual Travels Across Negative Scales},
  journal = tvcg,
  longjournal = {IEEE Transactions on Visualization and Computer Graphics},
  year = {2020},
  volume = {26},
  number = {1},
  pages = {654--664},
  doi = {10/kt3k},
  longdoi = {10.1109/TVCG.2019.2934334},
}

@inproceedings{Card:1991:TIV,
  author = {Card, Stuart K. and Robertson, George G. and Mackinlay, Jock D.},
  title = {The information visualizer, an information workspace},
  booktitle = {Proc.\ CHI},
  longbooktitle = {Proceedings of the SIGCHI Conference on Human Factors in Computing Systems},
  year = {1991},
  publisher = {ACM},
  address = {New York},
  optnumpages = {6},
  pages = {181--186},
  doi = {10/cvtdps},
  longdoi = {10.1145/108844.108874},
  opturl = {https://doi.org/10.1145/108844.108874},
  optlocation = {New Orleans, Louisiana, USA},
  optseries = {CHI '91},
}

@article{Cao:2025:FBV,
  author = {Zidi Cao and Jiayi Han and Sipeng Yang and Xiaogang Jin},
  title = {Fast best viewpoint selection with geometry-enhanced multiple views and cross-modal distillation},
  journal = vc,
  year = {2025},
  volume = {41},
  number = {7},
  pages = {5075--5086},
  doi = {10/p5p3},
  longdoi = {10.1007/s00371-024-03708-5},
}

@article{Yang:2019:DLB,
  author = {Changhe Yang and Yanda Li and Can Liu and Xiaoru Yuan},
  title = {Deep learning-based viewpoint recommendation in volume visualization},
  journal = jv,
  year = {2019},
  volume = {22},
  number = {5},
  pages = {991--1003},
  doi = {10/p5p4},
  longdoi = {10.1007/s12650-019-00583-4},
}

\clearpage

\begin{strip}
\noindent\begin{minipage}{\textwidth}
\makeatletter
\centering%
\sffamily\bfseries\fontsize{15}{16.5}\selectfont
\vgtc@title\\[.5em]
\large Supplementary Materials\\[.75em]
\makeatother
\normalfont\rmfamily\normalsize\noindent\raggedright We provide additional tables and charts beyond the material that we include in the main paper.
For access to the source code, datasets, and analysis scripts used in this work, please refer to \href{https://osf.io/hfu6e}{\texttt{osf.io/hfu6e}}. and to github repository \href{https://github.com/LixiangZhao98/ScaleFree}{\texttt{github.com\discretionary{/}{}{/}LixiangZhao98\discretionary{/}{}{/}ScaleFree}}.
\end{minipage}
\end{strip}

\appendix

\section{Workflow illustrations}
\label{appendix-flowchart}

\begin{figure}[t]
    \centering
    \includegraphics[width=\linewidth]{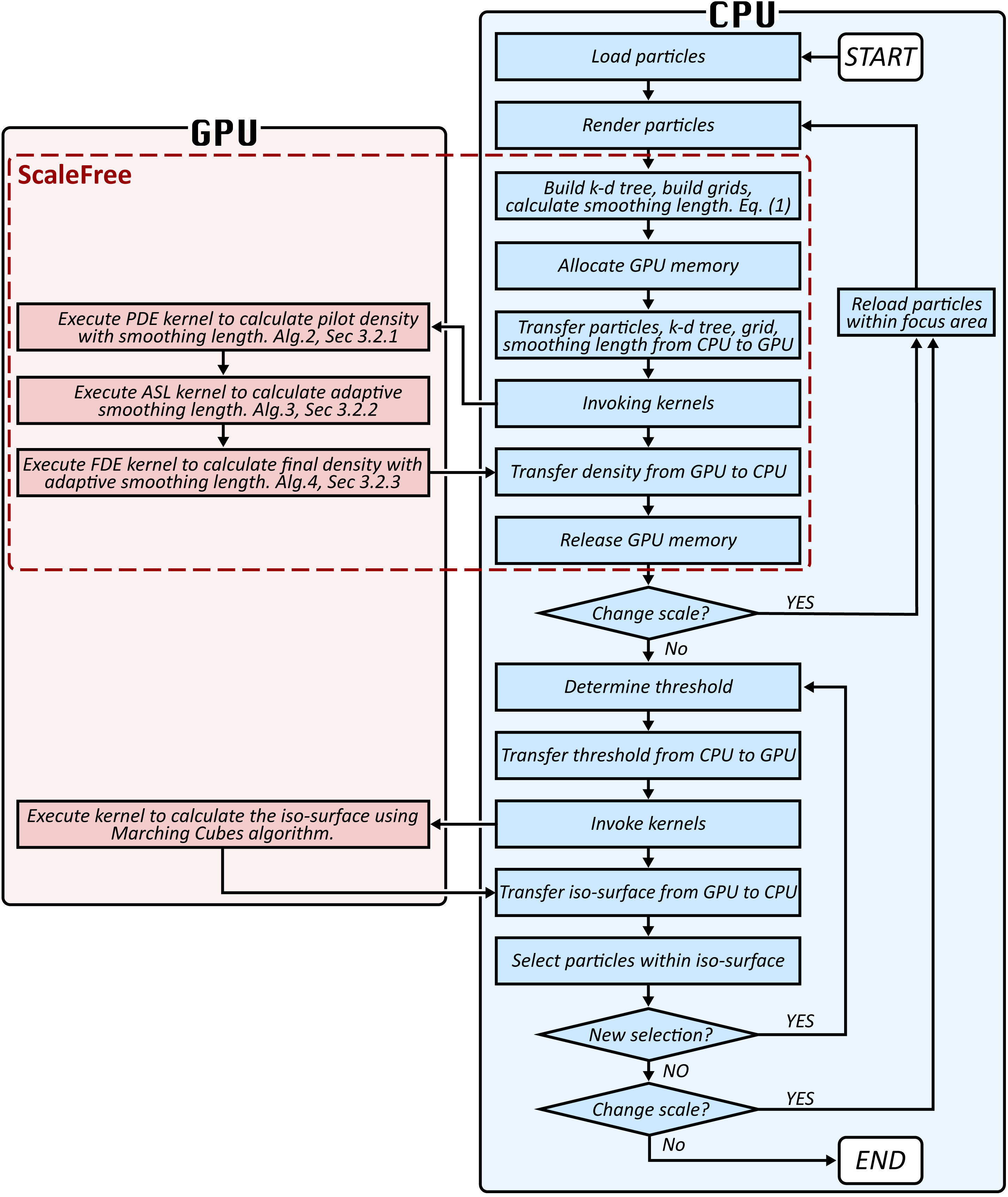}
    \caption{Workflow of adaptive selection method.}
    \label{fig:app:SelectionFlowchart}
\end{figure}

\begin{figure}[t]
    \centering
    \includegraphics[width=\linewidth]{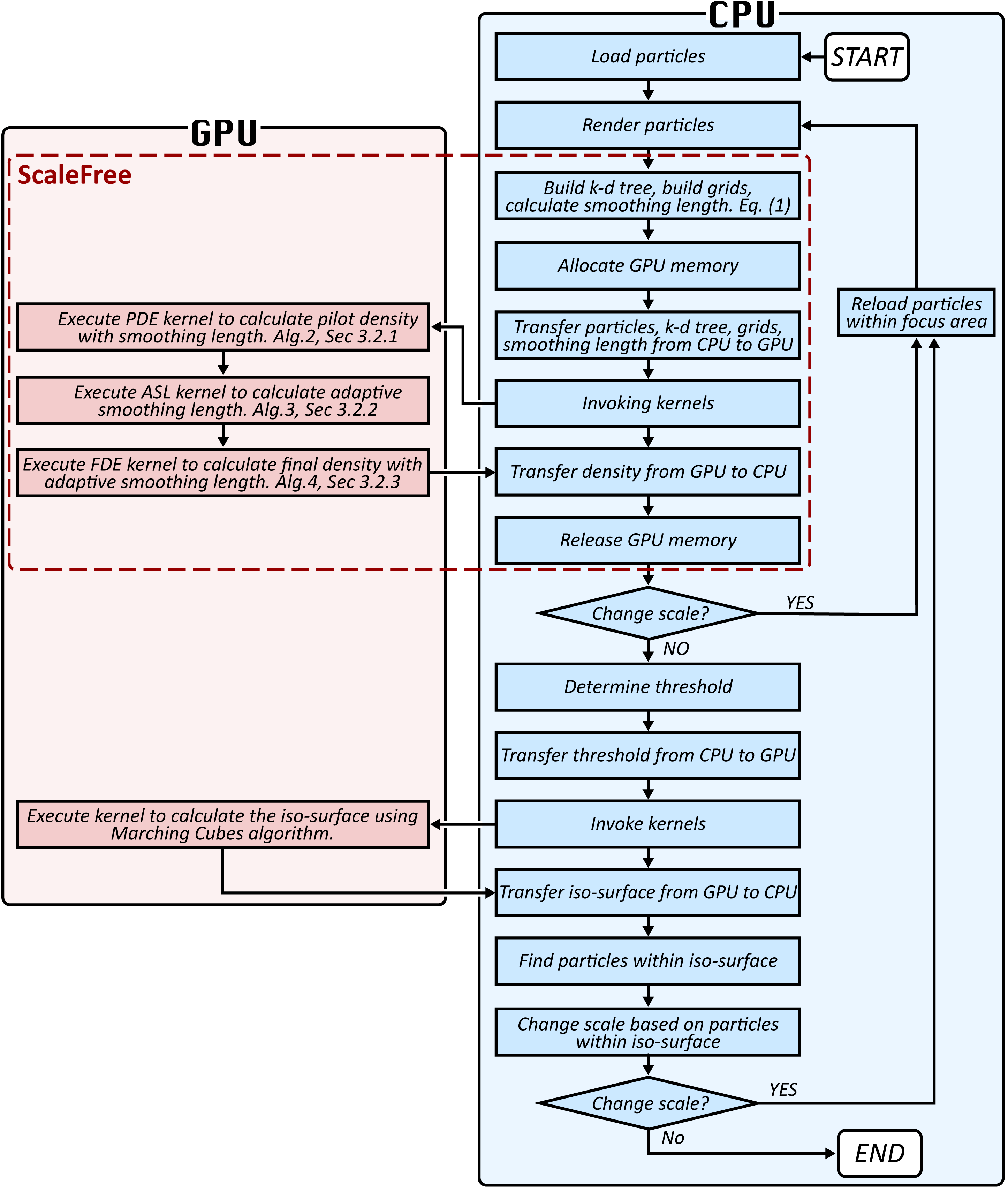}
    \caption{Workflow of progressive navigation method.}
    \label{fig:app:NavigationFlowchart}
\end{figure}

\begin{figure}[t]
    \centering
    \includegraphics[width=\linewidth]{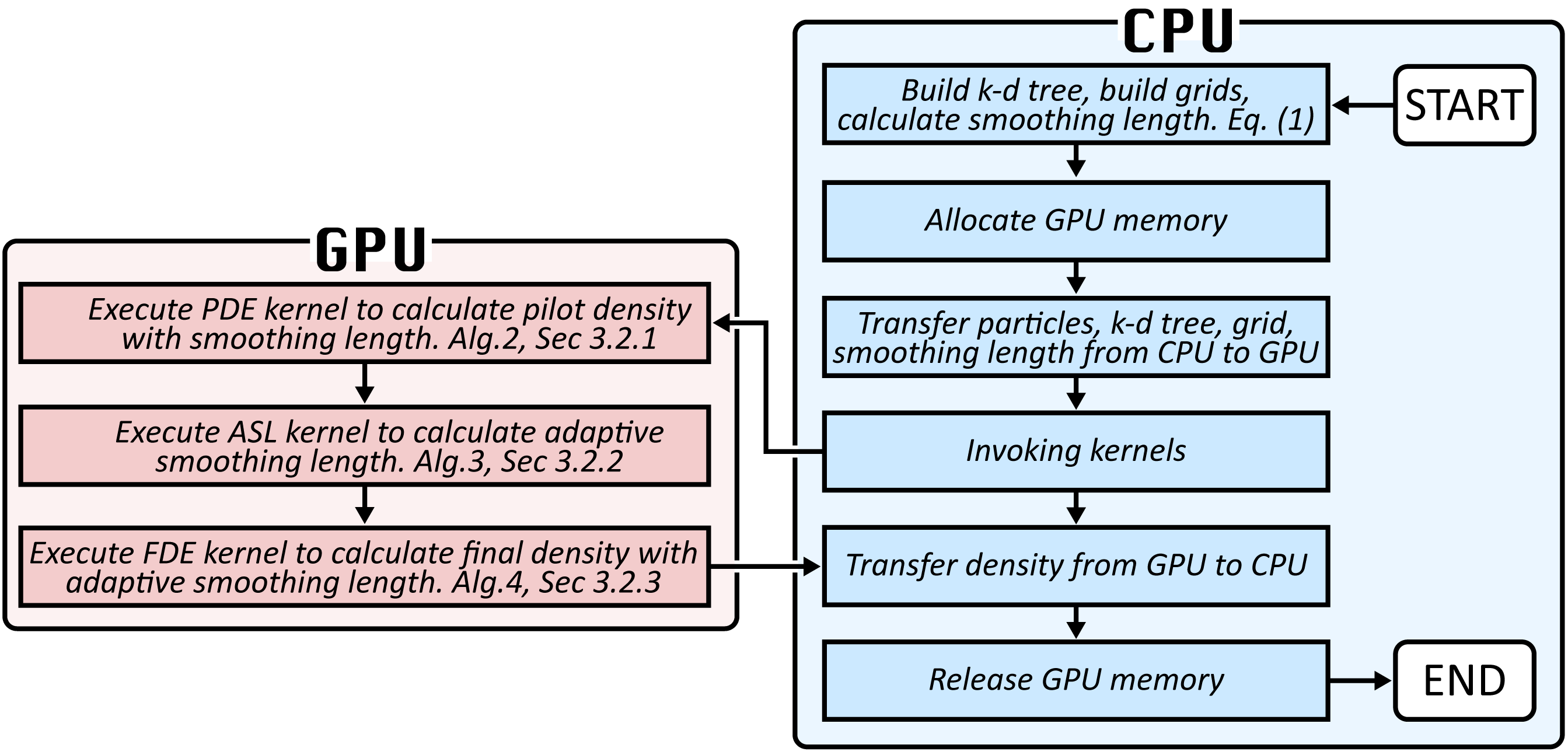}
    \caption{Workflow of the GPU-accelerated KDE algorithm.}
    \label{fig:app:flowchart}
\end{figure}

\autoref{fig:app:flowchart} shows the flow chart of our GPU-accelerated KDE algorithm described in \autoref{subsec:gpukde}. \autoref{fig:app:SelectionFlowchart} shows the flow chart of our scalable selection technique described in \autoref{subsec:selection:method}, while \autoref{fig:app:NavigationFlowchart} shows the flow chart of our progressive navigation technique described in \autoref{subsec:navigation:method}.

\section{Additional results from the study}
\label{appendix-result_userstudy}
\autoref{tab:app:timef1mcc} reports the mean completion times and accuracy scores, whereas \autoref{tab:app:pairwise} presents the pairwise ratios, as detailed in \autoref{subsec:Results}.
\begin{table}[b]
    \caption{The mean task completion times, accuracy scores, and their corresponding 95\% confidence intervals.}
    \label{tab:app:timef1mcc} 
    \centering
    \scriptsize%
    \centering%
    \begin{tabular*}{\hsize}{@{\extracolsep{\fill}}c | c c | c c | c c}
    \toprule
        Technique & Time & CI & FI & CI & MCC & CI \\ \midrule
        PS & 136s & [113,161] & .50 & [.44,.55] & .46 & [.41,.52] \\ 
        PM & 131s & [113,151] & .64 & [.59,.69] & .63 & [.58,.68] \\ 
        DR & 100s & [90,112] & .85 & [.79,.88] & .83 & [.78,.87] \\ \bottomrule
    \end{tabular*}
\end{table}

\begin{table}[b]
    \caption{The pairwise ratio of task completion times, accuracy scores, and their corresponding 95\% confidence intervals.}
    \label{tab:app:pairwise} 
    \centering
    \resizebox{\linewidth}{!}{%
        \begin{tabular}{c|cc|cc|cc}
            \toprule
            Technique & Time & CI & FI & CI & MCC & CI \\ \midrule
            DR/PM & 0.89 & [0.70,1.08] & 1.33 & [1.21,1.44] & 1.35 & [1.22,1.49] \\ 
            DR/PS & 0.86 & [0.66,1.08] & 1.74 & [1.58,1.92] & 1.90 & [1.67,2.18] \\ 
            PM/PS & 1.12 & [0.87,1.41] & 1.36 & [1.21,1.53] & 1.48 & [1.26,1.74] \\ 
            \bottomrule
        \end{tabular}%
    }
\end{table}


\section{Dataset used in user study}
\label{appendix-dataset_userstudy}
\begin{figure*}[p]
    \centering
    \includegraphics[width=\textwidth,height=\textheight,keepaspectratio]{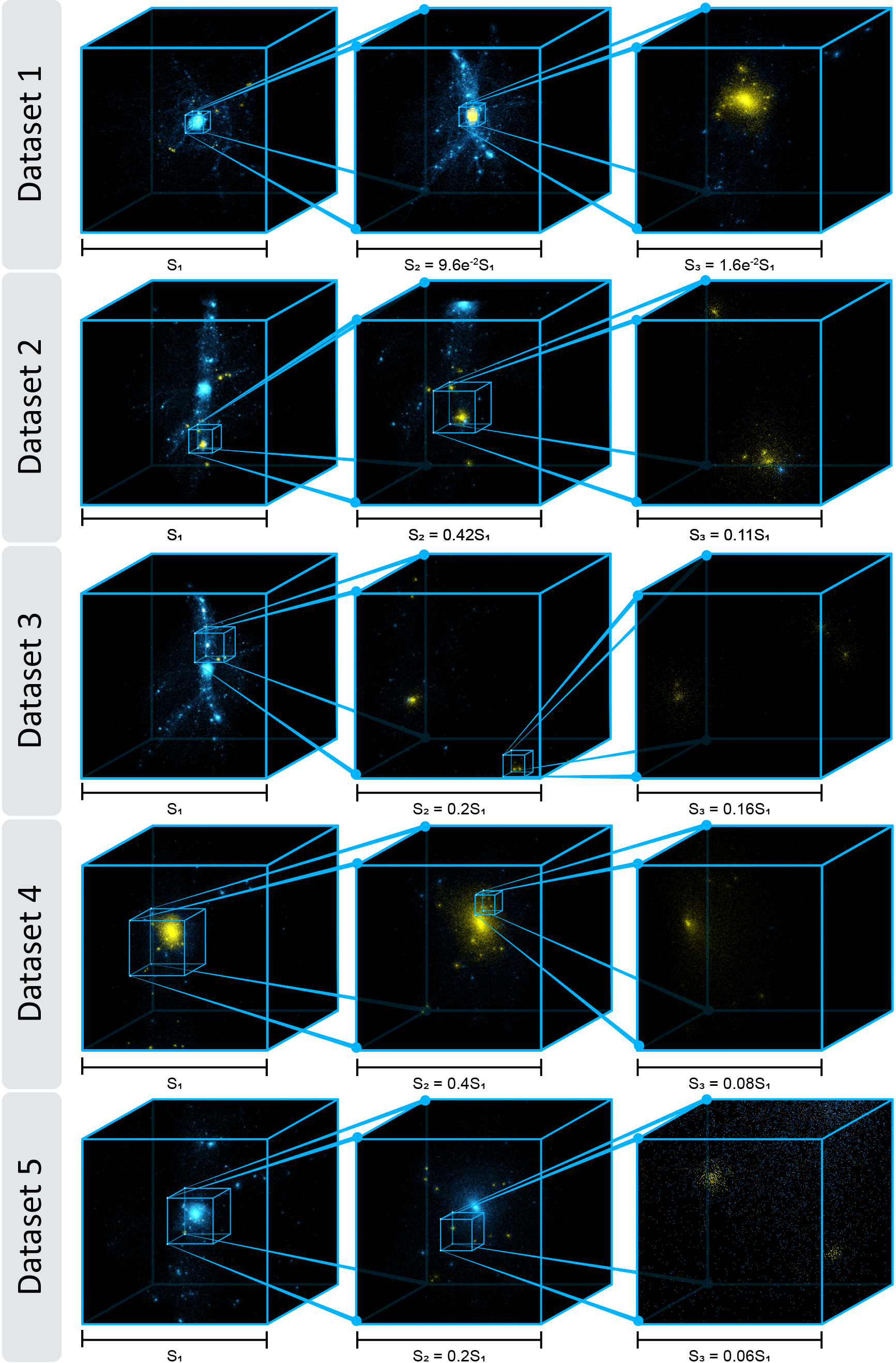}
    \caption{Experiment datasets.}
    \label{fig:app:datasets}
\end{figure*}

We extracted five timesteps from a cosmological N-body simulation~\cite{springel:2008:aquarius} and used them as our datasets. Tasks began at varying scales, with all targets becoming visible as participants zoomed in.
\autoref{fig:app:datasets} shows the five datasets that were used in the user study described in \autoref{sec:userstudy}.

\section{\lixiang{Parallel programming framework on the GPU}}
\label{appendix-gpuframe}

\lixiang{For the GPU parallel programming of our algorithm we follow the HLSL convention.
In this model, parallel computation is performed by invoking (programming) kernels, each of which is executed concurrently by multiple GPU threads. A thread is the basic unit of execution, running the same kernel function independently.
We can organize threads into groups defined by \texttt{numthreads(tx,\,ty,\,tz)} before launching a kernel, which specifies the dimension of threads in the group along the $x$-, $y$-, and $z$-dimensions.
Threads within a group share fast group memory and can synchronize their data exchange with group memory via the function \texttt{GroupMemoryBarrierWithGroupSync()}. Groups remain independent, thus, threads in different groups cannot access each other's group memory.
All GPU threads have access to global memory, which serves as the primary space for data exchange during computation. 
Before launching kernels, we need to transfer the input data from the CPU to global GPU memory. 
When launching a compute kernel, \texttt{Dispatch(kernel,\,groupCountX,\,groupCountY, groupCountZ)} specifies the number of thread groups along the $x$-, $y$-, and $z$-dimensions, thereby defining the overall dispatch layout.
The total number of threads executed equals the product of threads per group and the number of groups.
During kernel execution, each thread identifies its local (within its group), group, and global (among all threads) indices through semantics such as \texttt{SV\_GroupIndex}, \texttt{SV\_GroupID}, and \texttt{SV\_DispatchThreadID}.
Once the kernel completes execution, results are written to GPU buffers and can be returned to the CPU for further processing.}

\section{\lixiang{Interview questions}}
\label{appendix-inter_que}
\lixiang{During the interview phase of our user study, we asked our participants the following questions:}

\noindent\textbf{Q1. Perceived delay during interaction.}
\lixiang{Did you perceive any noticeable delays, lags, or interruptions during scale changes or while making selections when using each technique (PS, PM, and DR)?
If yes, please describe when the delay occurred (e.g., during scale transitions, selection, or navigation) and how noticeable it was.}

\noindent\textbf{Q2. Impact of delay on interaction experience.}
\lixiang{If you noticed any delays or performance slowdowns, did they affect your ability to navigate or select regions effectively?
Please explain your reasoning.}

\end{document}